\documentclass{jpp}
\usepackage{graphicx}
\usepackage{tensor}
\usepackage{bm}
\usepackage{enumitem}
\usepackage{stackrel}

\usepackage{tikz}
\newcommand*\circled[1]{\tikz[baseline=(char.base)]{
            \node[shape=circle,draw,inner sep=2pt] (char) {#1};}}

\newlist{myitemize}{itemize}{3}
\setlist[myitemize,1]{label=\textbullet,leftmargin=2em,rightmargin=2em,itemindent=0pt,labelsep=5pt,labelwidth=2em}
\setlist[myitemize,2]{label=$\rightarrow$,leftmargin=1em}
\setlist[myitemize,3]{label=$\diamond$}

\newlist{myenumerate}{enumerate}{1}
\setlist[myenumerate,1]{leftmargin=3em,rightmargin=3em,itemindent=0pt,labelsep=5pt,labelwidth=2em}

\usepackage[utf8]{inputenc}
\usepackage[T1]{fontenc}
\usepackage{amsmath}

\renewcommand{\tt}{\tilde{\theta}}							

\newcommand{\grad}{\boldsymbol{\nabla}}									

\makeatletter
\newcommand{\vast}{\bBigg@{3}}
\newcommand{\Vast}{\bBigg@{4}}
\makeatother

\title{The Second Quadratic Invariant of Astrophysical Gyrokinetics}

\author{B.~D.~G.~Chandran\aff{1}\footnote[1]{email:
    benjamin.chandran@unh.edu}, A.~Mallet\aff{2}, and R.~Meyrand\aff{1}}
\affiliation{\aff{1}Space Science Center and Department of Physics
  and Astronomy,  University of New Hampshire, Durham, NH 03824
\aff{2} Space Sciences Laboratory and Department of Physics, University of California, Berkeley, CA 94720
}

\begin{document}

\maketitle

\begin{abstract}
  We show that gyrokinetics in a spatially uniform equilibrium (or `astrophysical gyrokinetics') possesses a previously unknown quadratic invariant, which we call the gyrokinetic helicity.  We derive the limiting forms of the gyrokinetic helicity in five subsidiary expansions of astrophysical gyrokinetics: the isothermal electron fluid (ITEF) approximation, kinetic reduced magnetohydrodynamics (KRMHD), electron reduced magnetohydrodynamics (ERMHD), finite-Larmor-radius magnetohydrodynamics (FLR-MHD), and the kinetic reduced electron heating model~(KREHM). We subdivide the real part of the gyrokinetic helicity into `magnetofluid' and `phase-space' components, which are separately conserved in KRMHD, ERMHD, and FLR-MHD (in which there is no phase-space helicity). The magnetofluid helicity is equivalent to the Alfv\'enic part of the cross helicity in KRMHD, the magnetic helicity in~ERMHD, and the generalized helicity in FLR-MHD and~KREHM.  We derive an analytic expression for the rate at which magnetofluid helicity is converted into phase-space helicity at length scales comparable to the proton gyroradius in the ITEF approximation. We also explain how this conversion circumvents the helicity barrier and enhances turbulent heating in coronal holes and the near-Sun solar wind.
\end{abstract}

\section{Introduction}
\label{sec:intro} 

A quadratic invariant is a volume integral in space or phase space
with two properties: each term in the integrand contains two factors
of a fluctuating quantity, and the integral is conserved when
dissipation, forcing, and transport through the boundary of the
integration domain are neglected. 
Quadratic invariants play an important role in turbulence as they are
the quantities that are cascaded from large scales to small scales, or
vice versa, by nonlinear interactions. For example, gyrokinetics
possesses a quadratic invariant called the free energy, which is a linear
combination of the perturbations to the magnetic energy and plasma entropy. In gyrokinetic
turbulence, free energy is cascaded from large scales to small scales
in both velocity space and position space, leading to irreversible
collisional heating~\citep{schekochihin09}.  

When a system conserves two quadratic invariants associated with the
same fluctuating quantity or quantities, nonlinear interactions are in some
instances forced to transport the two invariants in opposite
directions in~$k$ space.  This occurs in 2D hydro (two-dimensional
hydrodynamic) turbulence, in which the energy
$\int_0^\infty E(k) \mathrm{d}k$ and enstrophy (mean-square vorticity)
$\int_0^\infty k^2 E(k) \mathrm{d}k$ are conserved. If (1)~energy and
enstrophy are injected into a system only at 
wavenumbers~$\sim k_0^{-1}$ and dissipate only at 
wavenumbers~$\sim k_{\rm d} \gg k_0$, (2)~the energy and enstrophy cascades are
dominated by local interactions in~$k$ space, and (3)~the rate at which energy
cascades to larger~$k$ in the inertial range~$k_0 \ll k\ll k_{\rm d}$
is $\propto k^0$, then
the rate at which enstrophy cascades to larger~$k$ must
be~$\propto k^2$, because the ratio of enstrophy to energy is~$k^2$. An
enstrophy cascade rate~$\propto k^2$ in the inertial range is
impossible, because nonlinear
interactions cannot generate enstrophy. This contradiction implies
that energy cannot cascade towards smaller scales in the inertial
range of 2D hydro turbulence \citep{fjortoft53}. Instead, it is only
the enstrophy that cascades to smaller scales, while the energy
cascades to larger scales. (We provide a more detailed discussion of
cascade directions in systems with multiple invariants in~\S\ref{sec:Fjortoft}.)

A related phenomenon occurs in finite-Larmor-radius MHD (FLR-MHD)
\citep{meyrand21}. FLR-MHD is a subsidiary expansion of gyrokinetics
that neglects non-Alfv\'enic fluctuations and assumes $\beta \ll 1$
and~$m_e \ll m_i$, where $\beta$ is the ratio of plasma to magnetic
pressure, and $m_e$ and~$m_i$ are the electron and ion masses. Like 2D
hydro, FLR-MHD possesses two quadratic invariants: the energy and
generalized helicity. In each of the two wavenumber ranges~$k_\perp
\rho_i \ll 1$ and $k_\perp \rho_i \gg 1$, 
these two invariants can be expressed as integrals
over~$k_\perp$ of the power spectra of the fluctuating quantities
multiplied by coefficients of the form~$c k_\perp^n$, where $k_\perp$
is the wavenumber perpendicular to the background magnetic field,
and $\rho_i$ is the ion gyroradius. At
$k_\perp \rho_i \ll 1$, the
value of~$n$ is the same for both invariants. However, at
$k_\perp \rho_i \gg 1$, $n$ is larger for the energy. Simultaneous
conservation of energy and generalized helicity causes the generalized
helicity to undergo an inverse cascade at~$k_\perp \rho_i \gg 1$
\citep{meyrand21}. If energy and generalized helicity are injected
only at $k_\perp \rho_i \ll 1$ and dissipate only at
$k_\perp \rho_i \gg 1$, then the inverse cascade of generalized
helicity at $k_\perp \rho_i \gg 1$ creates a `helicity barrier' that
prevents generalized helicity from reaching the small scales at which
it could dissipate, causing it to build up
secularly in time at $k_\perp \rho_i \ll 1$ \citep{meyrand21,
  squire22, squire23, adkins24, adkins25,zhang25}.

As these examples illustrate, quadratic invariants play a central role
in turbulence, and the existence of multiple quadratic invariants can
fundamentally alter how conserved quantities cascade. A complete
inventory of quadratic invariants is thus essential for understanding
plasma turbulence. In this paper, we point out that astrophysical
gyrokinetics --- that is, gyrokinetics in a
uniform plasma equilibrium ---
possesses a second three-dimensional quadratic invariant in addition to the free
energy.\footnote{We refer the reader to Appendix~F of
  \cite{schekochihin09} for a discussion of additional two-dimensional
  invariants that are conserved (in the absence of collisions,
  forcing, and boundary effects) when the fluctuating quantities are
  independent of the spatial coordinate along the background magnetic field.}
We call this invariant the gyrokinetic helicity~${\cal H}$.
In \S\ref{sec:GK}, we review the equations of astrophysical
gyrokinetics and use these equations to show that~${\cal H}$ is
conserved when collisions and transport through the plasma boundary
are negligible.  In \S\ref{sec:regimes}, we derive the limiting forms
of~${\cal H}$ in five subsidiary expansions of astrophysical gyrokinetics: the
isothermal electron fluid (ITEF) approximation; kinetic reduced MHD
(KRMHD); electron reduced MHD (ERMHD); FLR-MHD; and the kinetic
reduced electron heating model~(KREHM).  We subdivide the gyrokinetic
helicity in these subsidiary expansions into a `magnetofluid'
component~${\cal H}_{\rm mf}$ and a `phase-space'
component~${\cal H}_{\textrm{ph-sp}}$, show that ${\cal H}_{\rm mf}$
and~${\cal H}_{\textrm{ph-sp}}$ are separately conserved in certain
regimes, and derive the rate at which ${\cal H}_{\rm mf}$ transforms
into~${\cal H}_{\textrm{ph-sp}}$ at $k_\perp \rho_i \sim 1$ in the
ITEF approximation.  In \S\ref{sec:HB} we discuss the implications of
our findings for the helicity barrier and turbulent heating of coronal
holes (defined in \S\ref{sec:HB}) and the near-Sun solar wind.

\section{Quadratic invariants in astrophysical gyrokinetics}
\label{sec:GK}

Gyrokinetics was developed for toroidal plasmas and magnetic
confinement fusion by a number of authors
\citep[e.g.][]{rutherford68,taylor68,catto78,antonsen80,frieman82,abel13}. 
In this paper, we consider the simpler case of a uniform background
plasma (astrophysical gyrokinetics) and 
follow the approach of \cite{howes06} and \cite{schekochihin09}. We
assume that
\begin{equation}
  \frac{\delta n_s}{n_{0s}} \sim \frac{\delta T_s}{T_{0s}} \sim \frac{\delta B_\perp}{B_0}
  \sim\frac{\delta B_\parallel}{B_0} \sim \frac{u_\perp}{v_{\rm A}}
   \sim   \frac{k_\parallel}{k_\perp} \sim\frac{\omega}{\Omega_{\rm
       i}} \sim \epsilon_{\rm gk} \ll 1
  \label{eq:gkordering}
\end{equation}
and
\begin{equation}
 k_\perp\rho_i \sim \beta_{\rm s} \sim \frac{k_\parallel v_{\rm
     A}}{\omega} \sim  \frac{\nu_{\rm ii}}{\omega} \sim \mathrm{O}(1), 
  \label{eq:kperprho}
\end{equation}
where $n_{0s}$ ($\delta n_s$) is the average (fluctuating) number
density of species~$s$ ($s=i$ for ions and $s=e$ for electrons), $T_{0s}$ ($\delta T_s$) is the average
(fluctuating) temperature of species~$s$, $\bm{B}_0$ ($\delta \bm{B}$)
is the average (fluctuating) magnetic field, $\bm{B}_0 = B_0
\bm{\hat{z}}$, $\delta B_\perp$
($\delta B_\parallel$) is the component of~$\delta \bm{B}$
perpendicular (parallel) to~$\bm{B}_0$,
$u_\perp = c|\bm{E}\times \bm{B}_0|/B_0^2$, $\bm{E}$ is the electric
field, $v_{\rm A}$ is the Alfv\'en speed,
$k_\perp$ ($k_\parallel$) is the component of the
wavevector~$\bm{k}$ perpendicular (parallel) to~$\bm{B}_0$,
$\rho_s = v_{\rm th\it s}/\Omega_s$ is the gyroradius of species~$s$,
$v_{\rm th\it s} = \sqrt{2 T_{0s}/m_s}$
is the thermal speed of species~$s$, 
$\Omega_s= q_s B_0/m_s c$ is the ion cyclotron frequency of
species~$s$, $m_s$ ($q_s$) is the mass (charge) per particle of
species~$s$, $c$ is the speed of light,
and~$\epsilon_{\rm gk}$ is
the gyrokinetic ordering parameter.
In~(\ref{eq:kperprho}), $\beta_s = 8 \pi n_{0s} T_{0s} /B_0^2$, $\omega$ is the fluctuation
frequency, and~$\nu_{\rm ii}$ is the ion-ion Coulomb collision
frequency. 
Our assumption that $\omega \sim k_\parallel v_{\rm A}$ orders out
fast-magnetosonic-like fluctuations, for
which~$\omega \sim k_\perp v_{\rm A}$. We note that~(\ref{eq:kperprho})
does not restrict our analysis to values of
$k_\perp \rho_{\rm i}$,  $\beta_{\rm s}$,  $\omega/(k_\parallel
v_{\rm A})$, and $\nu_{\rm ii}/\omega$ close to~1. Rather, the
equations derived on the basis of~(\ref{eq:kperprho}) are uniformly valid
for values of these quantities greater or smaller than~1, and subsidiary orderings can
later be taken in which these quantities are $\ll 1$ or~$\gg 1$ (see
\S\ref{sec:regimes}).

The equations of gyrokinetics are obtained by expanding the
Vlasov-Landau equation and
Maxwell's
equations in powers
of~$\epsilon_{\rm gk}$ and solving order by order. The distribution
function of species~$s$ is written as
\begin{equation}
  f_s(\bm{r},\bm{v}, t) = F_{0s}(\bm{v},t) + \delta f_s(\bm{r}, \bm{v},t),
  \label{eq:deffs} 
\end{equation}
where $\bm{r}$ is position, $\bm{v}$ is velocity, and~$t$ is~time.
Coulomb collisions cause the zeroth-order distribution
function to be Maxwellian: $F_{0s} = n_{0s} \pi^{-3/2} v_{\rm th\it s}^{-3}
\exp(-v^2/v_{\rm th \it s}^2)$.  At first order
in~$\epsilon_{\rm gk}$, the perturbed distribution takes the form
\begin{equation}
  \delta f_s(\bm{r}, \bm{v}, t) = - \frac{q_s
    \phi(\bm{r},t)}{T_{0s}} F_{0s}(v) + h_s(\bm{R}_s, v_\perp, v_\parallel, t),
  \label{eq:dfs}
\end{equation}
where $\phi$ is the electrostatic potential, the term proportional
to~$\phi$ is the Boltzmann response,
\begin{equation}
  \bm{R}_{\rm s} = \bm{r} + \frac{\bm{v}_\perp \times
    \bm{\hat{z}}}{\Omega_{\rm s}}
  \label{eq:defRs}
\end{equation} 
is the particle guiding center or gyrocenter,
$v_\parallel = \bm{v}\cdot \bm{\hat{z}}$,
$\bm{v}_\perp = \bm{v} - v_\parallel \bm{\hat{z}} = \bm{\hat{x}}
v_\perp \cos\theta + \bm{\hat{y}}v_\perp \sin \theta$, and~$\theta$ is
the gyrophase angle.  The perturbed gyrocenter distribution
function~$h_s$ satisfies the gyrokinetic equation,
\begin{equation}
  \frac{\partial h_s}{\partial t} + v_\parallel \frac{\partial
    h_s}{\partial z} + \frac{c}{B_0} \left\{\langle
    \chi\rangle_{\bm{R}_s},h_s   \right\} = \frac{q_s }{T_{0s}}
  \frac{\partial \langle \chi\rangle_{\bm{R}_s}}{\partial t} F_{0s} +
  \left(\frac{\partial h_s}{\partial t}
    \right)_{\rm c},
  \label{eq:GKgeneral}
\end{equation}
where
\begin{equation}
  \chi = \phi -
  \frac{\bm{v} \cdot \bm{A}}{c}
  \label{eq:defchi}
\end{equation}
is the gyrokinetic potential, $\bm{A}$ is the magnetic vector
potential, $(\partial h_s/\partial
t)_{\rm c}$ is the rate of change of~$h_s$ due to Coulomb collisions,
\begin{equation}
\langle f(\bm{r}, \bm{v}, t) \rangle_{\bm{R}_s} \equiv
\frac{1}{2\pi}\int_0^{2\pi}
\mathrm{d}\theta\, f\left(\bm{R}_s - \frac{\bm{v}_\perp\times
    \bm{\hat{z}}}{\Omega_s},\bm{v},t\right)
\label{eq:gyroavR}
\end{equation}
is the gyroaverage of~$f$ at fixed~$\bm{R}_s$,
\begin{equation}
  \{f,g\} \equiv \bm{\hat{z}} \cdot \left(\grad f \times \grad g\right),
  \label{eq:poissonbracket}
\end{equation}
and $f$ and~$g$
in~(\ref{eq:gyroavR}) and~(\ref{eq:poissonbracket}) are arbitrary functions.
Throughout this paper, spatial derivatives are evaluated at
constant~$\bm{v}$, making the gradient operators with respect to~$\bm{r}$
and~$\bm{R}_s$ interchangeable. We assume that $v_{\rm th \it i} \ll
c$, which enables us to replace Poisson's equation with the
quasineutrality condition, 
\begin{equation}
\sum_s q_s \delta n_s = 
  \sum_s q_s \left( - \frac{q_s n_{0s}}{T_{0s}} \phi + \int \mathrm{d}^3
    \bm{v}\langle h_s\rangle_{\bm{r}}\right) = 0.
  \label{eq:QN}
\end{equation}
We adopt Coulomb gauge~$\grad \bcdot \bm{A} = 0$,
drop the (negligible) displacement current, and
write Ampere's Law in the form
\begin{equation}
  \sum_s q_s \int \mathrm{d}^3\bm{v} \langle
  h_s  \bm{v}\rangle_{\bm{r}} = - \frac{c}{4\pi} \nabla_\perp^2 \bm{A},
  \label{eq:Ampere} 
\end{equation} 
where $\grad_\perp \equiv \bm{\hat{x}} \frac{\partial}{\partial x} +
\bm{\hat{y}}\frac{\partial }{\partial y}$, and
\begin{equation}
\langle f(\bm{R}_s, \bm{v}, t) \rangle_{\bm{r}} \equiv
\frac{1}{2\pi}\int_0^{2\pi}
\mathrm{d}\theta\, f\left(\bm{r} + \frac{\bm{v}_\perp \times
    \bm{\hat{z}}}{\Omega_s},\bm{v},t\right)
\label{eq:gyroavr}
\end{equation}
is the gyroaverage of~$f$ at fixed~$\bm{r}$. 

In the absence of collisions, (\ref{eq:GKgeneral}),
(\ref{eq:QN}), and~(\ref{eq:Ampere}) conserve a quadratic invariant
known as the free energy \citep{schekochihin09},
\begin{equation}
  W = \int \frac{\mathrm{\rm d}^3 \bm{r}}{V} \left[\frac{|\delta
      \bm{B}|^2}{8\pi} + \sum_s\int \mathrm{d}^3 \bm{v} \frac{T_{0s} (\delta f_s)^2}{2F_{0s}}\right],
  \label{eq:WGK}
\end{equation}
where~$V$ is the plasma volume, which we take to be arbitrarily large.
The free energy is a linear combination of the perturbations to
the magnetic energy and plasma entropy~\citep{schekochihin09}
\begin{equation}
S = - \sum_s\int \frac{\mathrm{d}^3\bm{r}}{V}
    \int\mathrm{d}^3\bm{v} f_s \ln f_s =
     - \sum_s\int \frac{\mathrm{d}^3\bm{r}}{V}
    \int\mathrm{d}^3\bm{v} \left[F_{0s} \ln F_{0s} + \frac{(\delta
        f_s)^2}{2F_{0s}}\right] + \dots
    \label{eq:S}
\end{equation}

The central result of this paper is that, in the absence of collisions, (\ref{eq:GKgeneral}),
(\ref{eq:QN}), and~(\ref{eq:Ampere}) conserve a second quadratic
invariant that we call the gyrokinetic helicity,
\begin{equation}
  {\cal H} = \lim_{\epsilon \rightarrow 0^+} \sum_{s} T_{0s} \int
  \frac{\mathrm{d}^3\bm{R}_s}{V} \int \mathrm{d}^3\bm{v}
  \frac{\left(h_s - \displaystyle  q_s T_{0s}^{-1} \langle \chi
  \rangle_{\bm{R}_s} F_{0s} \right)^2}{2(v_\parallel - \mathrm{i} \epsilon v_{\rm th \it s})F_{0s}}.
  \label{eq:defHGK}
\end{equation}
The proof is straightforward. Neglecting the
$(\partial h_s/\partial t)_{\rm c}$ term in~(\ref{eq:GKgeneral}), we
obtain
\begin{align}
  \frac{\mathrm{d}{\cal H}}{\mathrm{d} t} & =
                                             \lim_{\epsilon \rightarrow 0^+} \sum_{s} T_{0s} \int
  \frac{\mathrm{d}^3\bm{R}_s}{V} \int \mathrm{d}^3\bm{v}
  \frac{\left(h_s - \displaystyle  q_s T_{0s}^{-1} \langle \chi
  \rangle_{\bm{R}_s} F_{0s} \right)}{(v_\parallel - \mathrm{i}
                                            \epsilon v_{\rm th \it
                                            s})F_{0s}} \left( -v_\parallel \frac{\partial
                                            h_s}{\partial z}\right) \nonumber \\
  & = - \sum_s \int
  \frac{\mathrm{d}^3\bm{R}_s}{V} \int\mathrm{d}^3\bm{v}\, q_s h_s
  \frac{\partial}{\partial z} \left\langle \phi - \frac{\bm{v} \cdot \bm{A}}{c}\right\rangle_{\bm{R}_s} \nonumber\\
&  =  - \int\frac{\mathrm{d}^3\bm{r}}{V} \left( \frac{\partial \phi}{\partial
    z} \sum_s q_s \int \mathrm{d}^3\bm{v} \langle
    h_s\rangle_{\bm{r}}  -
               \frac{1}{c}                            
 \frac{\partial \bm{A}}{\partial
    z}\cdot \sum_s  q_s\int \mathrm{d}^3\bm{v}  \langle h_s \bm{v}\rangle_{\bm{r}} \right) \nonumber\\
&  =  - \int\frac{\mathrm{d}^3\bm{r}}{V} \left( \frac{\partial \phi}{\partial
    z} \sum_s \frac{q_s^2 n_{0s}}{T_{0s}} \phi +
               \frac{1}{4\pi}                            
 \frac{\partial \bm{A}}{\partial
    z}\cdot \nabla_\perp^2 \bm{A}\right) \nonumber\\
                                          & =
                                            0. \label{eq:dHdtgen}
\end{align}
On the first line of~(\ref{eq:dHdtgen}), we dropped (boundary) terms of the form
$\int \mathrm{d}^3 \bm{R}_s V^{-1} a \{a,b\}$, as $V^{-1}a \{a,b \}$
is a total divergence (see~(\ref{eq:purediv})) whose integral becomes
negligible as~$V\rightarrow \infty$. To obtain the second line
of~(\ref{eq:dHdtgen}), we noted that
$ \lim_{V\rightarrow \infty} V^{-1} \int_{-\infty}^\infty
\mathrm{d}^3\bm{R}(\partial/\partial z) h_s^2/2 = 0$ and integrated by
parts. To obtain the third line, we changed integration variables
from~$\bm{R}_s$ to~$\bm{r}$. To obtain the fourth line, we
invoked~(\ref{eq:QN}) and~(\ref{eq:Ampere}). To arrive at the final
line, we integrated by parts to find that
$V^{-1}\int \mathrm{d}^3 \bm{r} (\partial \bm{A}/\partial z) \cdot
\nabla_\perp^2 \bm{A} = -(2V)^{-1}\sum_{i=1}^2 \sum_{j=1}^3 \int
\mathrm{d}^3\bm{r} (\partial/\partial z) (\partial A_j/\partial r_i)^2
= 0$, where $r_1\leftrightarrow x$ and~$r_2\leftrightarrow y$.

To write~${\cal H}$ in a form that is more transparently non-singular,
we separate~${\cal H}$ in~(\ref{eq:defHGK}) into real and imaginary parts
using the Plemelj relation, 
\begin{equation}
  {\cal H} = \sum_{s} \int
  \frac{\mathrm{d}^3\bm{R}_s}{V}  \int_0^\infty \mathrm{d}v_\perp  \pi 
  v_\perp  \left[
  \mathrm{P} \int_{-\infty}^\infty \frac{T_{0s}
    \psi_s^2}{ v_\parallel F_{0s}}\mathrm{d}v_\parallel
  + \mathrm{i} \pi \left(\frac{T_{0s}
    \psi_s^2}{F_{0s}}\right)_{v_\parallel = 0}\right],
\label{eq:realimag}
\end{equation}
where
\begin{equation}
  \psi_s \equiv h_s - \frac{q_s}{T_{0s}} \langle \chi
  \rangle_{\bm{R}_s} F_{0s},
  \label{eq:defpsi}
\end{equation}
and $\mathrm{P}$ indicates the Cauchy principal value of the
integral. Then, noting that $\mathrm{P}\int_{-\infty}^\infty \mathrm{d}v_\parallel\,
f(v_\parallel)/v_\parallel = 0$ if~$f(-v_\parallel) = f(v_\parallel)$,
we rewrite~(\ref{eq:realimag})  as
\begin{equation}
  {\cal H} = \sum_{s} \int
  \frac{\mathrm{d}^3\bm{R}_s}{V}  \int_0^\infty \mathrm{d}v_\perp  \pi
  v_\perp  \left[
   \int_{-\infty}^\infty \frac{2T_{0s}
    \psi_{\rm even} \psi_{\rm odd}}{ v_\parallel F_{0s}}\mathrm{d}v_\parallel
  + \mathrm{i} \pi \left(\frac{T_{0s}
    \psi_s^2}{F_{0s}}\right)_{v_\parallel = 0}\right],
\label{eq:defHGK2}
\end{equation}
where
\begin{equation}
  \psi_{\rm even}(\bm{R}_s,v_\perp, v_\parallel, t)= \frac{1}{2}
  \left[ \psi(\bm{R}_s,v_\perp, v_\parallel, t) +
\psi(\bm{R}_s,v_\perp, -v_\parallel, t) \right],
\label{eq:psieven}
\end{equation}
and
\begin{equation}
  \psi_{\rm odd}(\bm{R}_s,v_\perp, v_\parallel, t)= \frac{1}{2}
  \left[ \psi(\bm{R}_s,v_\perp, v_\parallel, t) -
\psi(\bm{R}_s,v_\perp, -v_\parallel, t) \right].
\label{eq:psiodd}
\end{equation}
The only factor of~$1/v_\parallel$ on the right-hand side of~(\ref{eq:defHGK2})
multiplies~$\psi_{\rm odd}$, and $\psi_{\rm odd}/v_\parallel$ remains finite as~$v_\parallel
\rightarrow 0$. \footnote{Collisions ensure that
 $\psi_s$ is continuous and differentiable at~$v_\parallel =0$, implying that $\psi_{\rm odd}
 \propto v_\parallel$ as $v_\parallel \rightarrow 0$.}

A quick alternative demonstration that~$\mathrm{Im}\, {\cal H}$
is conserved by the non-collisional terms in~(\ref{eq:GKgeneral}) follows
from evaluating these terms at~$v_\parallel= 0$, which yields
\begin{equation}
  \frac{\partial \psi_s}{\partial t}\Biggr|_{v_\parallel=0} +
  \frac{c}{B_0} \left\{\langle \chi\rangle_{\bm{R}_s},
    \psi_s\right\}\Biggr|_{v_\parallel = 0}
  = 0.
  \label{eq:GKvpar0}
\end{equation}
Equation~(\ref{eq:GKvpar0})  implies (via (\ref{eq:purediv})) that
\begin{equation}
\frac{\mathrm{d}}{\mathrm{d}t}  \int \frac{\mathrm{d}^3\bm{R}_s}{V} \left(\frac{T_{0s} \psi_s^2}{2
      F_{0s}}\right)_{v_\parallel = 0} = 0
  \label{eq:ImHcons}
\end{equation}
at each~$\bm{v}_\perp$, which immediately
yields~$(\mathrm{d}/\mathrm{d}t)\mathrm{Im}\,{\cal H} = 0$. We note
that one recovers the same conservation laws for the real and
imaginary parts of~${\cal H}$ whether one takes the limit
$\epsilon\rightarrow 0^+$ or~$\epsilon\rightarrow 0^-$ in~(\ref{eq:defHGK}).
In the remainder of this paper, we focus on the real part of~${\cal
  H}$ only.

\section{Quadratic invariants in subsidiary expansions of
  astrophysical gyrokinetics}
\label{sec:regimes}

\begin{figure}
\centerline{
  \includegraphics[width=8cm]{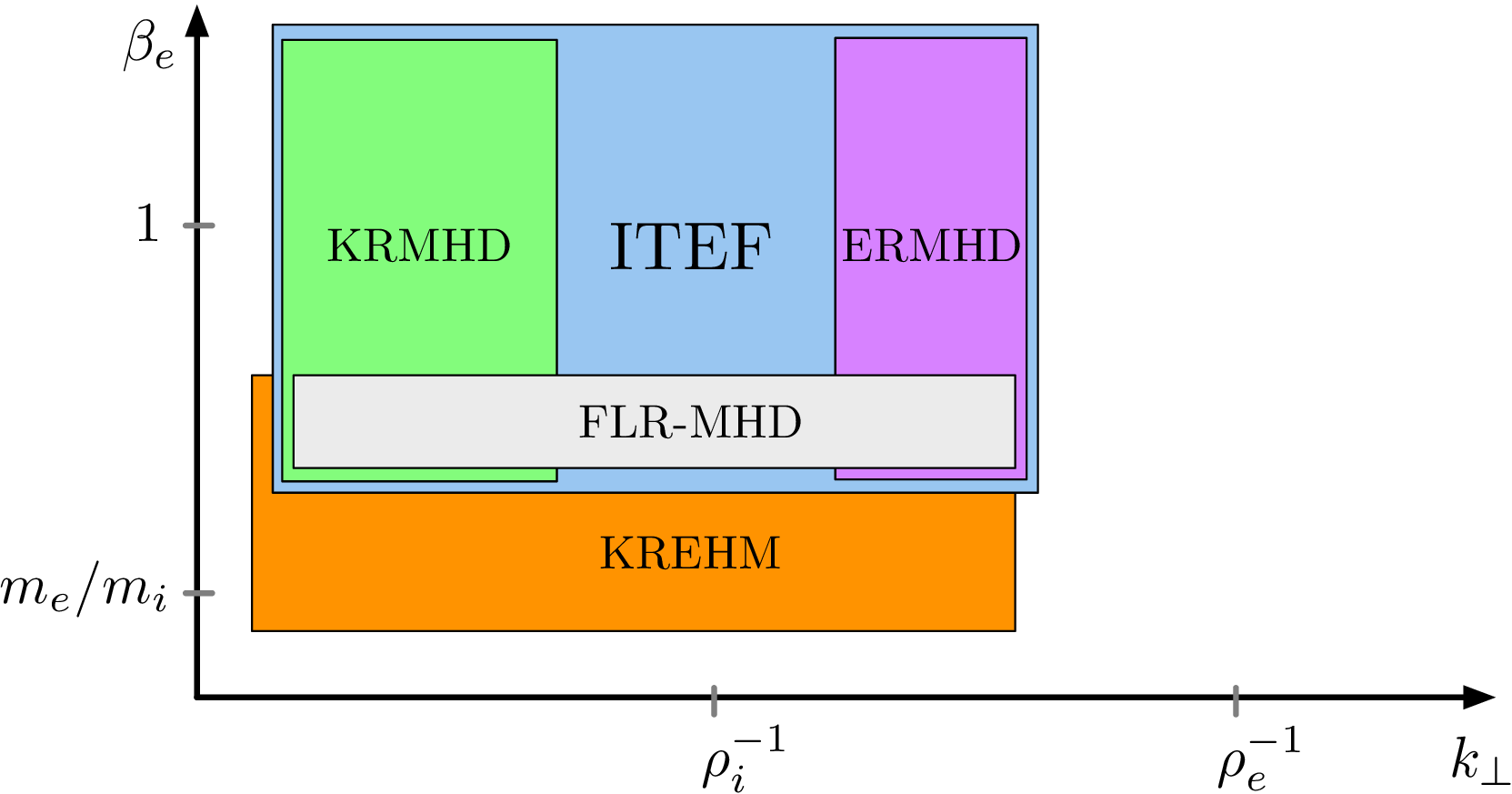}
  }
  \caption{Five subsidiary expansions of astrophysical gyrokinetics and their
    corresponding parameter regimes in the~$k_\perp - \beta_e$
    plane. These approximations to gyrokinetics are the
    isothermal electron fluid approximation, kinetic reduced
    magnetohydrodynamics, electron reduced
    magnetohydrodynamics, finite-Larmor-radius
    magnetohydrodynamics, and the kinetic reduced electron
    heating model.}
  \label{fig:regimes} 
\end{figure}

Well-known subsidiary expansions of the astrophysical gyrokinetic equations have
been carried out in the five parameter regimes illustrated in
figure~\ref{fig:regimes}. The ITEF approximation (regime acronyms are spelled out
in~\S\ref{sec:intro} and the caption of figure~\ref{fig:regimes}) sets
$k_\perp \rho_i \sim \mathrm{O}(1)$ and $\beta_e \sim \mathrm{O}(1)$
and expands the gyrokinetic equations to leading order
in~$(m_e/m_i)^{1/2}$. It allows $k_\perp \rho_i$ to be larger or
smaller than~1, but breaks down when $k_\perp \rho_i$ increases to
values $\gtrsim (m_i/m_e)^{1/2}$ or $\beta_e$ drops to values
$\lesssim m_e/m_i$.  KRMHD and ERMHD are subsidiary expansions of the
ITEF approximation in which $k_\perp \rho_i \ll 1$ and $k_\perp \rho_i \gg 1$,
respectively.  FLR-MHD is a $\beta_e \ll 1$  subsidiary expansion of
the ITEF approximation that neglects non-Alfv\'enic ion fluctuations. 
KREHM assumes that $\beta_e \sim m_e/m_i \ll 1$ and, like
FLR-MHD, neglects non-Alfv\'enic ion fluctuations. In all of these regimes,
\begin{equation}
  k_\perp \rho_e \ll 1, \qquad Z \equiv q_i/e \sim
  \mathrm{O}(1), \qquad \tau \equiv T_{0i}/T_{0e} \sim \mathrm{O}(1),
  \label{eq:kperprhoe}
\end{equation}
and it is assumed that there is a single ion species.
In the following subsections, we present the limiting forms of
the gyrokinetic helicity~${\cal H}$ in each of these subsidiary
regimes, subdivide~${\cal H}$ into `magnetofluid' and `phase-space'
components, and derive the rate at which~${\cal H}_{\rm mf}$ is converted
into~${\cal H}_{\textrm{ph-sp}}$ and vice versa.

\subsection{The isothermal electron fluid approximation}
\label{sec:ITEF}

The ITEF equations, which assume that $k_\perp \rho_e \ll 1$
and~$\beta_i \sim \beta_e \gg m_e/m_i$, were derived from (\ref{eq:GKgeneral}),
(\ref{eq:QN}), and~(\ref{eq:Ampere}) by \cite{schekochihin09} and
solved numerically by \cite{kawazura18} and \cite{kawazura19}.
Here, we present the ITEF equations using, for the most part, the compact notation of
\cite{schekochihin19}, in which
\begin{equation}
  \varphi = \frac{Z e \phi}{T_{0i }}= \frac{2\Phi}{\rho_{i}v_{\rm th\it i}}, \qquad A  = \frac{A_\parallel}{\rho_{i} B_0} = - \frac{\Psi}{\rho_{i}v_{\rm A}},
\label{eq:defphiA} 
\end{equation}
\begin{equation}
  \frac{\mathrm{d}}{\mathrm{d}t}\left(\dots\right) =
  \left(\frac{\partial}{\partial t}  + \bm{u}_\perp \cdot \grad_\perp\right)\left(\dots \right)
= \frac{\partial
  }{\partial t}(\dots) + \frac{\rho_{i } v_{\rm th\it i}}{2} \left\{ \varphi,
    \dots\right\},
  \label{eq:ddt}
\end{equation}
and
\begin{equation}
  \nabla_\parallel(\dots) =\left( \frac{\partial}{\partial z} +
  \frac{\bm{b}_\perp}{v_{\rm A}} \bcdot \grad_\perp\right)(\dots) = \frac{\partial}{\partial z}(\dots) - \rho_{i } \left\{ A, \dots\right\},
\label{eq:gradpar} 
\end{equation}
$ \bm{u}_\perp = \bm{\hat{z}} \times\grad_\perp \Phi = c \bm{E} \times
\bm{B}/B^2$,
$\bm{b}_\perp = \bm{\hat{z}}\times\grad_\perp \Psi = (\delta\bm{B} -
\bm{\hat{z}}\delta B)/\sqrt{4\pi n_{0i} m_i}$,
$\delta B = \bm{\hat{z}}\bcdot\delta \bm{B}$, $A_\parallel$~is the $z$
component of the vector potential, and
$v_{\rm A} = B_0/\sqrt{4\pi m_i n_{0i}}$. Following
\cite{schekochihin19}, we shall henceforth drop the~$0$ subscripts
on~$B_0$, $n_{0 s}$, and~$T_{0s}$ and the species subscripts on
$\delta n_i/n_i$ and~$\delta n_e/n_e$, as quasineutrality yields
\begin{equation}
  \frac{\delta n_i}{n_i} = \frac{\delta n_e}{n_e} \equiv \frac{\delta
    n}{n}.
  \label{eq:dnovern}
\end{equation}

In the ITEF approximation, there is no perturbation to the electron temperature,
$h_e$ equals~$F_{0e}(\bm{v})$ multiplied by a function
of~$\bm{R}_e$ (which is equivalent to $\bm{r}$ as $k_\perp \rho_e \ll
1$), and to leading order in $(m_e/m_i)^{1/2}$ the
electron gyrokinetic equation becomes
\begin{equation}
  \frac{\partial A}{\partial t}  = - \frac{v_{\rm th \it i}}{2} \nabla_\parallel \varphi + \frac{Zv_{\rm th \it i} }{2\tau}\nabla_\parallel \frac{\delta n}{n}
  \label{eq:dAdt}.
\end{equation} 
The electron continuity equation in the ITEF approximation is
\begin{equation} 
\frac{\mathrm{d}}{\mathrm{d}t} \left(\frac{\delta n}{n} - \frac{\delta
  B}{B}\right) + \nabla_\parallel u_{\parallel \rm e}  = -
\frac{\rho_{i } v_{\rm th \it i}}{2}\left\{ \frac{Z}{\tau} \frac{\delta
  n}{n} , \frac{\delta B}{B}\right\}. 
\label{eq:econt}
\end{equation} 

For the ions, we follow \cite{zocco11} and \cite{schekochihin19} and
rewrite~$\delta f_i$ as
\begin{equation}
\delta f_i(\bm{r},\bm{v},t) =\left[\left\langle
    \varphi\right\rangle_{\bm{R}_i(\bm{r},\bm{v}_\perp)} -
  \varphi(\bm{r},t)\right]F_{0i}(\bm{v}) + g_i\left(\bm{R}_i(\bm{r},\bm{v}_\perp),v_\perp,
v_\parallel,t\right), 
\label{eq:df}
\end{equation}
where
$\left(\left\langle \varphi\right\rangle_{\bm{R}_i} -
  \varphi\right)F_{0i}$ in~(\ref{eq:df}) is the
Alfv\'enic part of~$\delta f_i$, and $g_i$ is the non-Alfv\'enic
part. Following \cite{schekochihin19}, we define
\begin{equation}
  \overline{ (\dots)} = \frac{1}{n_i} \int \mathrm{d}^3 \bm{v}
  (\dots),
  \label{eq:overline}
\end{equation}
where the velocity integral in~(\ref{eq:overline}) is carried out at
fixed~$\bm{r}$. 
Applying this operator to~(\ref{eq:df}) yields
\begin{equation}
  \frac{\delta n}{n} = - \left(1 -\hat{\Gamma}_0\right) \varphi + \overline{\hat{J}_0 g_i}.
  \label{eq:dn}
\end{equation}
The differential operators~$\hat{J}_0$ and 
$\hat{\Gamma}_0$ appearing in~(\ref{eq:dn}), and the operators
$\hat{J}_1$ and~$\hat{\Gamma}_1$ appearing in the equations to follow,
arise from gyroaveraging\footnote{For example,
  $\langle p(\bm{R}_i)\rangle_{\bm{r}} =
  \hat{J}_0(a)p(\bm{r})$ for an arbitrary function~$p$.  This equation is
  obtained by expressing~$p$ as Fourier series
  $\sum p_{\bm{k}}e^{\mathrm{i}\bm{k}\cdot \bm{R}_i} = \sum
  p_{\bm{k}}e^{\mathrm{i}\bm{k}\cdot (\bm{r} + \bm{\rho})}$, 
gyroaveraging each term in
  the series, applying the identity
  $(2\pi )^{-1}\int_0^{2\pi} e^{ \mathrm{i} a \cos (\theta)} \mathrm{d}\theta =
  J_0(a)$, and replacing $J_0(k_\perp v_\perp/ \Omega_{i })$ with
  $J_0(-\mathrm{i}\hat{v}_\perp \hat{\nabla}_\perp)$.  Likewise,
  $\langle p(\bm{r}) \rangle_{\bm{R}_i} =
  \hat{J}_0(a)p(\bm{R}_i)$. These relations, combined with the identity
  $\int_0^\infty e^{-x^2} J_0(\lambda x)^2 x \mathrm{d}x = \frac{1}{2}
  e^{-\lambda^2/2}I_0(\lambda^2/2)$, imply that
  $ \overline{ \langle\langle
    \varphi(\bm{r})\rangle_{\bm{R}_i}\rangle_{\bm{r}}F_{0i}} =
  \hat{\Gamma}_0(\alpha) \varphi(\bm{r})$.}
and are given by~\citep{schekochihin19}
\begin{equation}
  \hat{J}_0(a) \leftrightarrow J_0(a) = 1 - \frac{a^2}{4} + \dots
  \qquad \hat{J}_1(a) \leftrightarrow \frac{2 J_1(a)}{a} = 1 -
  \frac{a^2}{8} \dots \qquad
  a^2 = - \hat{v}_\perp^2 \hat{\nabla}_\perp^2,
  \label{eq:J0J1}
\end{equation}
\begin{align}
  \hat{\Gamma}_0 \leftrightarrow \overline{ J_0^2(a) F_{0i}} =
  I_0(\alpha) e^{-\alpha} & = 1 - \alpha + \dots, \qquad \alpha = -
  \frac{1}{2} \hat{\nabla}_\perp^2, \\
\label{eq:Gamma0} 
   \hat{\Gamma}_1 \leftrightarrow \overline{ \hat{v}_\perp^2
                     \frac{2J_1(a) J_0(a)}{a}F_{0i}} & = - \frac{\mathrm{d}}{\mathrm{d}\alpha}\left[
                     I_0(\alpha) e^{-\alpha} \right] = 1 -
                     \frac{3}{2}\alpha + \dots ,
\end{align}
where  $\hat{v}_\perp = v_\perp/v_{\rm th \it i}$, $\hat{\grad}_\perp =\rho_{i } \grad_\perp$,
 $J_0$ and~$J_1$ are the ordinary Bessel functions of order~0 and~1,
respectively, and $I_0$ is the modified Bessel function of order~0.

Three additional equations close the system. The first two are
the equation of perpendicular pressure balance (or perpendicular
Ampere's law),
\begin{equation} 
  \frac{2}{\beta_{i }} \frac{\delta B}{B}  = - \frac{Z}{\tau} \frac{\delta n}{n} + \left(1 - \hat{\Gamma}_1\right) \varphi - \overline{\hat{v}_\perp^2 \hat{J}_1 g_i},
  \label{eq:dB} 
\end{equation} 
and the $z$ component of Ampere's law,
\begin{equation}
  \frac{u_{\parallel \rm e}}{v_{\rm th \it i}} = \frac{1}{\beta_{i }}
  \hat{\nabla}_\perp^2 A + \overline{ \hat{v}_\parallel \hat{J}_0 g_i},
  \label{eq:parAmpere}
\end{equation}
where $u_{\parallel \rm e}$ is the $z$ component of the mean electron
velocity,
 and~$\hat{v}_\parallel  = v_\parallel /v_{\rm th \it i}$. The final
 equation is the ITEF gyrokinetic equation for the perturbed ion distribution,  
\[
\frac{\partial }{\partial t} \left( g_i - \hat{v}_\perp^2 \hat{J}_1 \frac{\delta B}{B} F_{0i}\right) =
- \frac{\rho_{i }v_{\rm th \it i}}{2} \left( \left\{\langle \varphi\rangle_{\bm{R}_i}, g_i - \hat{v}_\perp^2 \hat{J}_1 \frac{\delta B}{B} F_{0i} \right\} +  \left\{\hat{v}_\perp^2 \hat{J}_1 \frac{\delta B}{B},g_i \right\}\right)
\]
\begin{equation}
  - v_\parallel \left\langle
    \nabla_\parallel \left(g_i + \frac{Z}{\tau}\frac{\delta n}{n} F_{0i}\right) +
    \rho_{i }\left\{A
      - \langle A \rangle_{\bm{R}_i} ,\varphi - \langle \varphi \rangle_{\bm{R}_i}\right\} F_{0i}
\right\rangle_{\bm{R}_i},
\label{eq:dgdt}
\end{equation}
where we have neglected the collisional term.

In the ITEF approximation, the gyrokinetic free energy in~(\ref{eq:WGK})  becomes
\[
W =  \frac{ n T_i}{2} \int \frac{\mathrm{d}^3\bm{r}}{V}\Biggr[
\frac{\overline{\langle g_i^2\rangle_{\bm{r}}}}{F_{0i}} + \varphi\left(
    1-\hat{\Gamma}_0\right)\varphi
  + \frac{Z}{\tau} \left|\left(1- \hat{\Gamma}_0\right)\varphi -
    \overline{ \hat{J}_0 g_i} \right|^2
\nonumber
\]
\begin{equation} 
  + \frac{2}{\beta_{i }} \left(|\hat{\nabla}_\perp A|^2+
     \frac{\delta\!B^2}{B^2}\right)
 \Biggr],
 \label{eq:W}
\end{equation}
and the real part of the gyrokinetic helicity in~(\ref{eq:defHGK}) becomes
\begin{equation}
  \mathrm{Re} \,{\cal H} = {\cal H}_{\rm mf} + {\cal H}_{\rm \textrm{ph-sp} },
  \label{eq:H1H2}
\end{equation}
where
\begin{equation} 
{\cal H}_{\rm mf}  = m_i n_i v_{\rm th\it i}\int \frac{\mathrm{d}^3 \bm{r}}{V} A \left[ \left(1 -
             \hat{\Gamma}_0\right) \varphi + \left(1 -
             \hat{\Gamma}_1\right) \frac{\delta
             B}{B}\right] \label{eq:H1},
       \end{equation}
and
\begin{equation}        
{\cal H}_{\rm \textrm{ph-sp} }  =  T_i\int \frac{\mathrm{d}^3
  \bm{R}_i}{V} \int_0^\infty \mathrm{d}v_\perp \pi v_\perp \mathrm{P}
\int_{-\infty}^\infty \mathrm{d}v_\parallel 
\frac{\left(g_i - \hat{v}_\perp^2 \hat{J}_1 \displaystyle \frac{\delta
      B}{B} F_{0i} \right)^2}{v_\parallel F_{0i} }  .
  \label{eq:H2}
\end{equation}
We employ the the subscript~`mf,' short for `magnetofluid,'
in~(\ref{eq:H1}) because the right-hand side
of~(\ref{eq:H1}) contains the magnetic
fluctuations~($A$, $\delta B$) and low-order moments of the ion
distribution function via~(\ref{eq:dn}).
In~(\ref{eq:H2}), we employ the subscript~`ph-sp,' short for
`phase-space,' because the right-hand side
of~(\ref{eq:H2}) contains contributions from the small-scale
velocity-space structure of~$g_i$. For example, if~$g_i$ is expanded
in Hermite polynomials as in Appendix~\ref{ap:phasemixing}, then all the
Hermite coefficients of~$g_i$ contribute to~${\cal  H}_{\textrm{ph-sp}}$.

We show in  \S\ref{ap:dHdt} that, when collisions are neglected,
\begin{align}
  \frac{\mathrm{d}}{\mathrm{d}t} &{\cal H}_{\rm mf} =  - n_i T_i
                                             \int\frac{\mathrm{d}^3\bm{r}}{V}\frac{Z}{\tau}
  \left(\hat{\Gamma}_1 \frac{\delta B}{B}\right) \nabla_\parallel
                                             \frac{\delta n}{n}
                                                \nonumber \\
  &  - n_i T_i
    \int\frac{\mathrm{d}^3\bm{r}}{V} \int\frac{\mathrm{d}^3 \bm{v}}{n_i}
    \Biggr\langle \left[\frac{Z}{\tau} \frac{\delta n(\bm{r},t)}{n} +
    \hat{v}_\perp^2 \hat{J}_1 \frac{\delta B(\bm{R}_i,t)}{B}
    \right]\nabla_\parallel g_i(\bm{R}_i,v_\perp,v_\parallel,t) \nonumber \\
  & + \rho_i \left[g_i(\bm{R}_i,v_\perp,v_\parallel,t) - \hat{v}_\perp^2
    \hat{J}_1 \frac{\delta B(\bm{R}_i,t)}{B} F_{0i}   \right]
    \Big[ \big\{ \varphi({\bm{r}},t),A(\bm{r},t) \big\}  -
                 \big\{\langle \varphi\rangle_{\bm{R}_i},\langle
    A\rangle_{\bm{R}_i} \big\} \Big]
    \Biggr\rangle_{\bm{r}},
    \label{eq:dH1dt0}
\end{align}
where, on the second line, the~$A$ appearing in $\nabla_\parallel
g_i(\bm{R}_i, v_\perp, v_\parallel, t)$ via (\ref{eq:gradpar})
is evaluated at~$\bm{r}$. We also verify, using~(\ref{eq:dAdt})
through~(\ref{eq:dgdt}), that
\begin{equation}
  \frac{\mathrm{d}}{\mathrm{d}t}{\cal H}_{\rm \textrm{ph-sp} } = - \frac{\mathrm{d}}{\mathrm{d}t}{\cal
    H}_{\rm mf},
\label{eq:dH2dt0}
\end{equation} 
as required by~(\ref{eq:dHdtgen}) and~(\ref{eq:H1H2}).
The conversion of ${\cal H}_{\rm mf}$ to~${\cal H}_{\textrm{ph-sp}}$
may play an important role
in circumventing the helicity barrier in coronal holes and the
near-Sun solar wind, enabling more turbulent heating than would be
expected on the basis of FLR-MHD, a point that we discuss in greater
detail in~\S\ref{sec:HB}. At the end of Appendix~\ref{ap:dHdt}, we
describe how the leading-order terms on the right-hand side
of~(\ref{eq:dH1dt0}) cancel at $k_\perp \rho_i \ll 1$ and become small
 because of gyroaveraging at~$k_\perp \rho_i \gg 1$, consistent with
our findings in \S\ref{sec:KRMHD} and~\S\ref{sec:ERMHD}  that
${\cal H}_{\rm mf}$ and~${\cal H}_{\textrm{ph-sp}}$ are separately
conserved in KRMHD and~ERMHD. Significant conversion
of~${\cal H}_{\rm mf}$ to~${\cal H}_{\textrm{ph-sp}}$ (or vice versa)
thus occurs only at $k_\perp \rho_i \sim 1$.

\subsection{Kinetic reduced magnetohydrodynamics}
\label{sec:KRMHD}

\cite{schekochihin09} derived the equations of KRMHD by 
expanding the ITEF equations in powers of $k_\perp \rho_i \ll 1$. As shown by these authors, 
to satisfy~(\ref{eq:gkordering})
when~$k_\perp \rho_i \ll 1$, $\varphi$ and~$A$ become~$\sim
(k_\perp \rho_i)^{-1}$ times larger than~$\delta n/ n$ and~$\delta
B/B$. As a consequence, the electron
gyrokinetic equation~(\ref{eq:dAdt}) simplifies to the reduced
magnetohydrodynamic (RMHD) induction equation,
\begin{equation}
  \frac{\partial A}{\partial t} + \frac{v_{\rm th\it i}}{2}
  \nabla_\parallel \varphi = 0,
  \label{eq:gkeKRMHD}
\end{equation}
which implies that the electric field
along the total magnetic field vanishes.
Subtracting the ion continuity equation from the electron continuity
equation yields the RMHD vorticity equation,
\begin{equation}
  \frac{1}{2} \frac{\mathrm{d}}{\mathrm{d}t} \nabla_\perp^2 \varphi +
  \frac{v_{\rm th \it i}}{\beta_i} \nabla_\parallel \nabla_\perp^2 A =
  0.
  \label{eq:vorticity}
\end{equation}
The (collisionless) ion gyrokinetic equation~(\ref{eq:dgdt}) simplifies to
\begin{equation}
  \frac{\mathrm{d}g_{09}}{\mathrm{d}t} + v_\parallel \nabla_\parallel\left[g_{09} + \left(\frac{Z}{\tau} \frac{\delta n}{n} +
      \hat{v}_\perp^2 \frac{\delta B}{B} \right) F_{0i}\right]
  = 0,
  \label{eq:KRMHDdgdt}
\end{equation}
where
\begin{equation}
  g_{09} \equiv g_i - \hat{v}_\perp^2 \frac{\delta B}{B} F_{0i}
  \label{eq:defg09}
\end{equation}
is the definition of~$g$ from~\cite{schekochihin09}, and
$g_i$ (see~(\ref{eq:df})) is the definition of~$g$
in~\cite{schekochihin19}.
In KRMHD, (\ref{eq:dn}) reduces to
\begin{equation}
\overline{ g_{09}} = \frac{\delta n}{n} -
\frac{\delta B}{B},
\label{eq:dnKRMHD}
\end{equation}
and the perpendicular part of Ampere's law (which yields perpendicular
pressure balance) reduces to
\begin{equation}
  \overline{ \hat{v}_\perp^2 g_{09}} = - \frac{Z}{\tau}\frac{\delta
    n}{n} - 2\left(1 + \frac{1}{\beta_i}\right) \frac{\delta B}{B}.
  \label{eq:perpAmpereKRMHD} 
\end{equation} 
Equations~(\ref{eq:gkeKRMHD}), (\ref{eq:vorticity}), (\ref{eq:KRMHDdgdt}), 
(\ref{eq:dnKRMHD}), and~(\ref{eq:perpAmpereKRMHD}) provide a complete model for the time
evolution of~$A$, $\varphi$, $g_{09}$,
$\delta n/n$, and $\delta B/B$.

In KRMHD, the free energy in (\ref{eq:W}) reduces to~\citep{schekochihin09}
\begin{equation}
  W = W_{\rm AW} + W_{\rm compr},
  \label{eq:WKRMHD}
\end{equation}
where
\begin{equation}
  W_{\rm AW} = \frac{n_i T_i}{2}\int \frac{\mathrm{d}^3\bm{r}}{V}
  \left( \frac{1}{2} |\hat{\grad}_\perp \varphi|^2 + \frac{2}{\beta_i}
    |\hat{\grad}_\perp A|^2\right)
  \label{eq:KRMHDWAW}
\end{equation}
and
\begin{equation}
  W_{\rm compr} = \frac{n_i T_i}{2} \int \frac{\mathrm{d}^3\bm{r}}{V}
  \left[
\frac{Z}{\tau} \left(\frac{\delta n}{n}\right)^2 + \frac{2}{\beta_i}
\left(\frac{\delta B}{B}\right)^2 + \int\frac{\mathrm{d}^3\bm{v}}{n_i} \frac{g_i^2}{F_{0i}} \right],
    \label{eq:KRMHDWcompr}
  \end{equation}
where the subscripts~`AW' and~`compr' stand for `Alfv\'en wave' and
`compressive,' respectively.
These two components of the free energy are separately conserved in
the absence of collisions~\citep{schekochihin09}. Likewise, in KRMHD,
the two components of $\mathrm{Re}\,{\cal H}$ in~(\ref{eq:H1H2})
reduce to
\begin{equation}
  {\cal H}_{\rm mf} = - \frac{1}{2} m_i n_i v_{\rm th\it i} \int
  \frac{\mathrm{d}^3\bm{r}}{V} A \hat{\nabla}_\perp^2 \varphi = -
  \frac{m_i n_i}{B} \int \frac{\mathrm{d}^3 \bm{r}}{V} \bm{u}_\perp
  \bcdot \delta \bm{B}_\perp,
  \label{eq:HAKRMHD} 
\end{equation}
which is proportional to the RMHD cross helicity, and
\begin{equation}        
{\cal H}_{\rm \textrm{ph-sp} }  =  T_i\int \frac{\mathrm{d}^3
  \bm{r}}{V} \int_0^\infty \mathrm{d}v_\perp \pi v_\perp \mathrm{P}
\int_{-\infty}^\infty \mathrm{d}v_\parallel 
\frac{g_{09}^2}{v_\parallel F_{0i} }  .
  \label{eq:HnAKRMHD} 
\end{equation}
It follows  from (\ref{eq:gkeKRMHD})
through~(\ref{eq:perpAmpereKRMHD})  that, in the absence of
collisions and when~$V\rightarrow \infty$, ${\cal H}_{\rm mf}$
and~${\cal H}_{\rm \textrm{ph-sp} }$ are separately conserved:
\begin{equation}
  \frac{\mathrm{d}}{\mathrm{d}t}{\cal H}_{\rm mf} = 0, \qquad
  \frac{\mathrm{d}}{\mathrm{d}t}{\cal H}_{\rm \textrm{ph-sp} } = 0.
  \label{eq:HAHnAcons}
\end{equation}

In Appendix~\ref{ap:phasemixing}, we review the finding of~\cite{schekochihin09}
that~$W_{\rm compr}$ further subdivides into three separately
conserved energies and show that ${\cal H}_{\rm \textrm{ph-sp} }$  likewise
subdivides into three separately conserved helicities. We also show
that KRMHD possesses two additional quadratic invariants  and discuss
their implications for parallel phase
mixing in imbalanced KRMHD turbulence.

\subsection{Electron reduced magnetohydrodynamics}
\label{sec:ERMHD} 

\cite{schekochihin09} derived the ERMHD equations from the ITEF equations
in the limit that $k_\perp \rho_i \gg 1$. In
this limit, the terms containing
$\hat{J}_0$, $\hat{J}_1$, $\hat{\Gamma}_0$,
and~$\hat{\Gamma}_1$ in (\ref{eq:dn}), (\ref{eq:dB}),  and~(\ref{eq:parAmpere}) 
can be neglected, and these equations become, respectively,
\begin{equation}
  \frac{\delta n}{n} = - \varphi, \qquad
\frac{\delta B}{B}  = \frac{\beta_i}{2}\left(1 + \frac{Z}{\tau}\right)
\varphi, \qquad
  \frac{u_{\parallel \rm e}}{v_{\rm th \it i}} = \frac{1}{\beta_{i }}
  \hat{\nabla}_\perp^2 A.
  \label{eq:dndBuERMHD}
\end{equation}
The leading-order electron gyrokinetic equation~(\ref{eq:dAdt}) then simplifies to
\begin{equation}
\frac{\partial A}{\partial t} = - \frac{v_{\rm th \it i}}{2}\left(1 +
  \frac{Z}{\tau}\right)\nabla_\parallel \varphi,
\label{eq:gkeERMHD} 
\end{equation}
the electron continuity equation becomes
\begin{equation}
  \frac{\partial \varphi}{\partial t} = \frac{\beta_i^{-1/2} v_{\rm
      A}}{1 + (\beta_i/2)(1 + Z/\tau)} \nabla_\parallel
  \hat{\nabla}_\perp^2 A,
  \label{eq:econtERMHD}
\end{equation}
and the (collisionless) ion gyrokinetic equation becomes
\begin{equation}
  \frac{\partial h_i}{\partial t} + v_\parallel \frac{\partial
    h_i}{\partial t} + \frac{\rho_i v_{\rm th \it i}}{2} \{\langle
  \varphi \rangle_{\bm{R}_i}, h_i\} = 0.
  \label{eq:gkiERMHD}
\end{equation}
Equation~(\ref{eq:gkiERMHD})is equivalent to equation~(249) of~\cite{schekochihin09}
when the collisional term and subdominant $\partial \langle \varphi
\rangle_{\bm{R}_i}/\partial t$ term are neglected.

In ERMHD, the free energy reduces to~\citep{schekochihin09}
\begin{equation}
  W = W_{\rm KAW} + W_{h_i},
  \label{eq:WERMHD}
\end{equation}
where
\begin{equation}
  W_{\rm KAW} = \frac{n_i
      T_i}{2} \int \frac{\mathrm{d}^3 \bm{r}}{V}\left\{
    \frac{2}{\beta_i}|\hat{\nabla}A|^2 + \left(1+\frac{Z}{\tau}\right)
    \left[1 + \frac{\beta_i}{2}\left(1 +
        \frac{Z}{\tau}\right)\right]\varphi^2\right\},
  \label{eq:WKAW}
\end{equation}
the subscript KAW stands for kinetic Alfv\'en wave, and
\begin{equation}
  W_{h_i} = \int \frac{\mathrm{d}^3\bm{R}_i}{V} \int \mathrm{d}^3\bm{v}
  \frac{T_{i} h_i^2}{2F_{0i}}.
  \label{eq:Whi}
\end{equation}
When collisions and boundary effects are neglected,
$W_{h_i}$ and~$W_{\rm KAW}$ are
separately conserved.
The two components of $\mathrm{Re}\,{\cal H}$ reduce to
\begin{equation}
 {\cal H}_{\rm mf} = \left[1 + \frac{\beta_i}{2}\left(1 + \frac{Z}{\tau}\right)\right]m_i n_i v_{\rm th\it i} \int \frac{\mathrm{d}^3
   \bm{r}}{V} A \varphi ,
  \label{eq:HAERMHD} 
\end{equation}
which is proportional to the magnetic helicity~\citep{schekochihin09}
as~$\varphi \propto \delta B/B$, and
\begin{equation}
  {\cal H}_{\rm \textrm{ph-sp} } = T_i
  \int\frac{\mathrm{d}^3\bm{R}_i}{V} \int_0^\infty \mathrm{d}v_\perp
  \pi v_\perp \mathrm{P}\int_{-\infty}^\infty \mathrm{d}v_\parallel
  \frac{h_i^2}{v_\parallel F_{0i}}.
  \label{eq:HnAERMHD} 
\end{equation}
Equations~(\ref{eq:gkeERMHD}), (\ref{eq:econtERMHD}), 
and~(\ref{eq:gkiERMHD}) imply  that~${\cal H}_{\rm mf}$ and~${\cal H}_{\textrm{ph-sp}}$ are separately conserved in ERMHD when collisions and boundary
effects are neglected:
\begin{equation}
  \frac{\mathrm{d}}{\mathrm{d}t}{\cal H}_{\rm mf} = 0, \qquad
  \frac{\mathrm{d}}{\mathrm{d}t} {\cal H}_{\rm \textrm{ph-sp} }= 0.
  \label{eq:HAHnAconsERMHD}
\end{equation}
The ERMHD (collisionless) conservation
law~$(\mathrm{d}/\mathrm{d}t)\int \mathrm{d}^3\bm{r} A \varphi = 0$,
i.e., the first equation in~(\ref{eq:HAHnAconsERMHD}),
was previously obtained
by~\cite{schekochihin09}.

\subsection{Finite-Larmor-radius magnetohydrodynamics}
\label{sec:FLRMHD}

FLR-MHD is a subsidiary expansion of the ITEF approximation in which $\beta \ll 1$
(implying that~$\delta B/B$ is negligible), and in which~$g_i$ is for
simplicity set equal to~0 \citep{meyrand21}.  In FLR-MHD, the
leading-order electron gyrokinetic equation retains the same form as
in the ITEF approximation,
\begin{equation}
  \frac{\partial A}{\partial t}  = - \frac{v_{\rm th \it i}}{2} \nabla_\parallel \varphi + \frac{Zv_{\rm th \it i} }{2\tau}\nabla_\parallel \frac{\delta n}{n}
  \label{eq:dAdtFLRMHD},
\end{equation} 
the electron continuity equation~(\ref{eq:econt}) reduces to
\begin{equation} 
\frac{\mathrm{d}}{\mathrm{d}t} \frac{\delta n}{n} + \nabla_\parallel u_{\parallel \rm e}  = 0,
\label{eq:econtFLRMHD}
\end{equation} 
the parallel component of Ampere's law (\ref{eq:parAmpere}) simplifies to
\begin{equation}
  \frac{u_{\parallel \rm e}}{v_{\rm th \it i}} = \frac{1}{\beta_{i }}
  \hat{\nabla}_\perp^2 A, 
  \label{eq:parAmpereFLRMHD}
\end{equation}
and  (\ref{eq:dn}) becomes
\begin{equation}
  \frac{\delta n}{n} = - \left(1 -\hat{\Gamma}_0\right) \varphi.
  \label{eq:dnFLRMHD}
\end{equation}
The free energy energy~(\ref{eq:W}) in FLR-MHD is 
\begin{equation} 
W =  \frac{ n T_i}{2} \int \frac{\mathrm{d}^3\bm{r}}{V}\Biggr[
\varphi\left(
    1-\hat{\Gamma}_0\right)\varphi
  + \frac{Z}{\tau} \left|\left(1- \hat{\Gamma}_0\right)\varphi\right|^2
  + \frac{2}{\beta_{i }}|\hat{\nabla}_\perp A|^2
 \Biggr],
 \label{eq:WFLRMHD}
\end{equation}
and the two components of $\mathrm{Re}\,{\cal H}$ are
\begin{equation} 
{\cal H}_{\rm mf}  = m_i n_i v_{\rm th\it i}\int \frac{\mathrm{d}^3 \bm{r}}{V} A\left(1 -
  \hat{\Gamma}_0\right) \varphi,
\qquad {\cal H}_{\rm \textrm{ph-sp} } = 0.
\label{eq:H1FLRMHD}
         \end{equation}
The magnetofluid helicity~${\cal H}_{\rm mf}$ in FLR-MHD is
proportional to the `generalized helicity' of FLR-MHD, which
is proportional to the RMHD cross helicity~(\ref{eq:HAKRMHD})  at~$k_\perp
\rho_i \ll 1$ and the magnetic helicity~(\ref{eq:HAERMHD}) at~$k_\perp \rho_i \gg 1$ \citep{meyrand21}.
Equations~(\ref{eq:dAdtFLRMHD}) through (\ref{eq:dnFLRMHD})  imply
that~\citep{meyrand21}
\begin{equation}
  \frac{\mathrm{d}}{\mathrm{d}t} {\cal H}_{\rm mf}= 0.
  \label{eq:dHmfdtFLRMHD}
\end{equation}

\subsection{The kinetic reduced electron heating model}
\label{sec:KREHM} 

KREHM is a low-$\beta$, small-mass-ratio expansion of the gyrokinetic
equations, first obtained by \cite{zocco11}, in which 
\begin{equation}
  \beta_{\rm e} \sim \beta_{\rm i} \sim \frac{m_e}{m_i} \ll 1,
\qquad
k_\perp \rho_e \ll 1.
\label{eq:paramKREHM} 
\end{equation}
To facilitate contact with \cite{zocco11}'s
presentation, we write the equations in this subsection
in terms of~$A_\parallel$ and~$\phi$ rather than the dimensionless~$A$
and~$\varphi$ used in \S\ref{sec:ITEF} through~\S\ref{sec:FLRMHD}.
The second equality in~(\ref{eq:paramKREHM}) implies that the electron
gyroradius~$\bm{R}_e$ and position~$\bm{r}$ are interchangeable and
that electron gyroaverages can be ignored.
As in FLR-MHD, the perturbation to the ion distribution function in
KREHM is taken to be purely Alfv\'enic:
\begin{equation}
  \delta f_i(\bm{r},\bm{v},t) =\frac{Ze}{T_{\rm i}}\left[\left\langle
    \phi\right\rangle_{\bm{R}_i(\bm{r},\bm{v}_\perp)} -
  \phi(\bm{r},t)\right]F_{0i}(\bm{v}).
\label{eq:dfiKREHM}
\end{equation}
Applying the $\overline{ (\dots)}$ operator of~(\ref{eq:overline})
to~(\ref{eq:dfiKREHM})  yields
\begin{equation}
  \frac{\delta n}{n} = - \frac{Ze}{T_{\rm i}}\left(1 -\hat{\Gamma}_0\right) \phi,
  \label{eq:dnKREHM}
\end{equation}
which, by quasineutrality, is the fractional density fluctuation of
both the ions and electrons. \cite{zocco11} wrote the perturbed
electron gyrocenter distribution function in the form
\begin{equation}
  h_e = \left(-\frac{e\phi}{T_{\rm e}} + \frac{\delta n}{n} + \frac{2v_\parallel u_{\parallel e}}{v_{\rm th \it e}^2} \right)F_{0e} +
  g_e,
  \label{eq:he}
\end{equation}
where
\begin{equation}
  \int \mathrm{d}^3 \bm{v} \left( \begin{array}{c} 1 \\
                                    v_\parallel \end{array}\right) g_e
                                = 0.
                                \label{eq:gemoments}
                              \end{equation}
That is, $g_e$ contains all information about electron temperature fluctuations, heat-flux
fluctuations, and all higher $v_\parallel$ moments of~$h_e$, but $g_e$
contributes to neither $\delta n/n$ nor~$u_{\parallel \rm e}$.
Given~(\ref{eq:dfiKREHM}), the parallel ion current vanishes, so the
parallel component of Ampere's law becomes
\begin{equation}
u_{\parallel \rm e} = \frac{e d_{\rm e}^2}{m_{\rm e}c} \nabla_\perp^2 A_\parallel,
  \label{eq:upareKREHM}
\end{equation}
where $d_{\rm e} = c/\omega_{\rm p\it e}$ is
the electron inertial length and $\omega_{\rm p \it e} = (4\pi n_e
e^2/m_e)^{1/2}$ is the electron plasma frequency.
The electron continuity equation in KREHM is then
\begin{equation}
  \frac{\partial }{\partial t} \frac{\delta n}{n} = -
  \frac{c}{B}\left\{\phi, \frac{\delta n}{n}\right\} -
  \nabla_\parallel \frac{ed_{\rm e}^2}{m_{\rm e}c} \nabla_\perp^2
  A_\parallel,
  \label{eq:econtKREHM}
\end{equation}
and the parallel component of the electron momentum equation is
\begin{equation}
  \frac{\partial }{\partial t}\left( A_\parallel - d_{\rm e}^2
    \nabla_\perp^2 A_\parallel\right) = - \frac{c}{B}\left\{\phi,
    A_\parallel - d_{\rm e}^2 \nabla_\perp^2 A_\parallel \right\} - c
  \frac{ \partial \phi}{\partial z} + \frac{cT_{\rm e}}{e}
  \nabla_\parallel \frac{\delta n}{n} + \frac{c}{e}\nabla_\parallel
  \delta T_{\parallel e}.
  \label{eq:parmomentumKREHM} 
\end{equation}
By subtracting (\ref{eq:econtKREHM}) multiplied by~$F_{0e}$ from
the electron gyrokinetic equation~(\ref{eq:GKgeneral}) and
using~(\ref{eq:parmomentumKREHM}) to eliminate~$\partial A_\parallel
/\partial t$, \cite{zocco11} obtained an equation for the time
evolution of~$g_e$, which (neglecting collisions) can be written in
the form
\begin{equation}
  \frac{\partial }{\partial t}g_e = -
  \frac{c}{B_0}\left\{\phi,g_e\right\} - v_\parallel \nabla_\parallel
  \left(g_e - \frac{\delta T_{\parallel e}}{T_e}F_{0e}\right)
  + \left(1 - \frac{2 v_\parallel^2}{v_{\rm th
        e}^2}\right)F_{0e}\nabla_\parallel \frac{e d_e^2}{m_e
    c}\nabla_\perp^2 A_\parallel.
  \label{eq:dgdtKREHM} 
\end{equation}
As $\delta T_{\parallel \rm e}$ can be computed directly from~$g_{\rm
  e}$, the five equations (\ref{eq:dnKREHM}), (\ref{eq:upareKREHM}),
(\ref{eq:econtKREHM}), (\ref{eq:parmomentumKREHM}),
and~(\ref{eq:dgdtKREHM})  determine the time evolution of
the five unknowns~$\delta n/n$, $u_{\parallel e}$, $\phi$,
$A_\parallel$, and~$g_e$.

In KREHM, the free energy~(\ref{eq:WGK}) is given by
\[
  W = \int\frac{\mathrm{d}^3\bm{r}}{V} \Biggr\{\frac{1}{8\pi}\left|\nabla_\perp
    A_\parallel\right|^2  + \frac{d_e^2}{8\pi} \left(\nabla_\perp^2 A_\parallel\right)^2
  + \frac{n_i T_i}{2} \frac{Ze\phi}{T_i} \left(1 -
    \hat{\Gamma}_0\right)\frac{Ze\phi}{T_i}
\]
\begin{equation} 
  + \frac{n_e T_e}{2} \left[\left(1 - \hat{\Gamma}_0\right) \frac{Ze\phi}{T_i}\right]^2
+ \int \mathrm{d}^3\bm{v} \frac{T_e g_e^2}{2 F_{0e}}\Biggr\},
  \label{eq:WKREHM}
\end{equation}
and $\mathrm{Re}\,{\cal H}$ reduces to
\begin{equation}
\mathrm{Re}\,  {\cal H} = {\cal H}_{\rm mf, KREHM} + {\cal H}_{\textrm{ph-sp},\rm KREHM},
\label{eq:HKREHM} 
\end{equation}
where
\begin{equation}
  {\cal H}_{\rm mf, KREHM} = \frac{e}{c}
  \int\frac{\mathrm{d}^3\bm{r}}{V} \delta n_e \left(d_e^2
    \nabla_\perp^2 A_\parallel - A_\parallel\right),
  \label{eq:HmfKREHM}
\end{equation}
and
\begin{equation}
  {\cal H}_{\textrm{ph-sp},\rm KREHM} = T_e \int\frac{\mathrm{d}^3
    \bm{r}}{V} \int_0^\infty \mathrm{d}v_\perp \pi v_\perp \mathrm{P}
  \int_{-\infty}^\infty \mathrm{d}v_\parallel
  \frac{\left(g_e+\displaystyle 
    \frac{\delta n}{n}F_{0e}\right)^2}{v_\parallel F_{0e}}   .
  \label{eq:HkinKREHM}
\end{equation}
The generalized helicity found by \cite{adkins25} is proportional
to~${\cal H}_{\rm mf, KREHM}$. With the aid of~(\ref{eq:dnKREHM}),
(\ref{eq:econtKREHM}), and~(\ref{eq:parmomentumKREHM}), we obtain
\begin{equation}
  \frac{\mathrm{d} }{\mathrm{d}t}{\cal H}_{\rm mf,KREHM}= - \int
  \frac{\mathrm{d}^3 \bm{r}}{V} \delta n_e \nabla_\parallel
  \delta T_{\parallel e},
  \label{eq:dHmfdtKREHM}
\end{equation}
which is equivalent to~(26) and~(27)
of~\cite{adkins25} in the absence of hyperviscosity and forcing. It
follows from (\ref{eq:econtKREHM}) and~(\ref{eq:dgdtKREHM}) that
\begin{equation}
  \frac{\mathrm{d} }{\mathrm{d}t}{\cal H}_{\textrm{ph-sp},\rm KREHM} = 
\int
  \frac{\mathrm{d}^3 \bm{r}}{V} \delta n_e \nabla_\parallel
  \delta T_{\parallel e},
\label{eq:dHentrdtKREHM} 
  \end{equation}
as required by conservation of~${\cal H}$ and~(\ref{eq:dHmfdtKREHM}).

\section{The helicity barrier in the isothermal electron fluid approximation}
\label{sec:HB}

In \S\ref{sec:intro}, we described how the inverse cascade of
generalized helicity at $k_\perp \rho_i \gg 1$ in FLR-MHD creates a
`helicity barrier' that prevents generalized helicity injected at
$k_\perp \rho_i \ll 1$ from reaching~$k_\perp \rho_i \gg 1$.  In this
section, we describe a cascade channel that circumvents the helicity barrier
in the more general ITEF approximation, which allows for
non-Alfv\'enic fluctuations and a larger range of~$\beta_{\rm e}$
values. 

\subsection{Cascade directions of simultaneously conserved invariants}
\label{sec:Fjortoft}

To predict and understand cascade directions in astrophysical
gyrokinetics, we draw upon \cite{alexakis18}'s generalization of
\cite{fjortoft53}'s analysis to account for sign-indefinite
invariants, which we paraphrase in terms of cascade rates rather
than dissipation rates. We consider two invariants~A and~B that
involve the same fluctuating quantities and have spectra~$E_{\rm A}$
and~$E_{\rm B}$. We assume that the cascade rates of~A and~B,
denoted~$\epsilon_{\rm A}$ and~$\epsilon_{\rm B}$, are dominated by local
interactions in~$k$ space and that the inertial range is very
broad. If there exist positive constants~$c$ and~$n$ such that
\begin{equation}
  E_{\rm B} \geq c k^n |E_{\rm A}|,
  \label{eq:AB1}
\end{equation}
then A cannot cascade to larger~$k$ at a
rate~$\epsilon_{\rm A}\propto k^0$ in the inertial range, for
then~$\epsilon_{\rm B}$ would have to increase with~$k$ at least as fast
as~$k^n$. Likewise, if there exist positive constants~$c$ and~$n$ for
which
\begin{equation}
  E_{\rm B} \geq c k^{-n} |E_{\rm A}|,
  \label{eq:AB2}
\end{equation}
then~A cannot undergo an inverse
cascade, for then the inverse cascade rate of~B would have to
increase with decreasing~$k$ at least as fast as~$k^{-n}$. 
The wavenumber~$k$ in (\ref{eq:AB1}) and~(\ref{eq:AB2}) is generic and
does not restrict the argument to isotropic~3D turbulence. We
apply the same arguments to the anisotropic cascade in wavenumber space in gyrokinetics, in which~$k$ is
replaced by~$k_\perp$, and to parallel phase mixing in velocity space, in which~$k$ is
replaced by the Hermite number~$m$ (see~Appendix~\ref{ap:phasemixing}).

In 2D hydro, the enstrophy spectrum is exactly~$k^2$ times the energy
spectrum, and thus (\ref{eq:AB1}) is satisfied with~B being the enstrophy
and~A being the energy, implying that energy cannot cascade to
larger~$k$. This recovers the result of \cite{fjortoft53} presented in~\S\ref{sec:intro}.
Equation~(\ref{eq:AB2})  is also satisfied in 2D~hydro, but this time
with B the energy and~A the enstrophy, implying that enstrophy
cannot undergo an inverse cascade. 

In 3D (three-dimensional) hydro, the
energy~$\int |\bm{u}|^2 \mathrm{d}^3 \bm{x}$ and
helicity~$\int \bm{u} \cdot \bm{\omega} \mathrm{d}^3\bm{x}$ are
conserved, where~$\bm{\omega} = \grad \times \bm{u}$ is the
vorticity. The helicity spectrum is thus the energy spectrum
multiplied by~$k \cos \theta_k$ where~$\theta_k$ is the angle between
the Fourier transforms of~$\bm{u}$ and~$\bm{\omega}$. Because the
invariant ``with more powers of~$k$'' (namely the helicity) is sign
indefinite (by virtue of the factor of~$\cos\theta_k$), (\ref{eq:AB1})
is not satisfied, and no constraint prevents energy from cascading to
larger~$k$. We note that \cite{milanese21} has shown that $\cos \theta_k$
decreases as~$1/k$ in 3D hydro turbulence, enabling constant fluxes of
both energy and helicity to larger~$k$ in the inertial range.
Although~(\ref{eq:AB1})  is not satisfied in 3D~hydro, (\ref{eq:AB2})~is, with A
being the energy and~B being the helicity, and thus an inverse
cascade of helicity is prohibited.

In ERMHD and FLR-MHD at $k_\perp \rho_i \gg 1$, the ratio of
KAW energy  to magnetofluid helicity ${\cal H}_{\rm mf}$ 
 is a constant times $k_\perp / \sigma_k$, where
$\sigma_k = (|\Theta^+_k|^2- |\Theta_k^-|^2)/(|\Theta^+_k|^2 +
|\Theta^-_k|^2)$, and $\Theta^+_k$ and~$\Theta^-_k$ are the amplitudes
of KAWs traveling parallel and antiparallel
to~$\bm{B}$ \citep{schekochihin09,meyrand21}.
Because $-1 \leq \sigma_k \leq 1$, (\ref{eq:AB1}) is satisfied
with~B being the KAW energy and~A being~${\cal H}_{\rm mf}$,
implying that ${\cal H}_{\rm mf}$ cannot cascade to larger~$k_\perp$.
The reason that (\ref{eq:AB1}) prohibits a forward cascade in ERMHD
and FLR-MHD at~$k_\perp \rho_i \gg 1$ but not in 3D~hydro, even though
all three cases involve a sign-indefinite invariant, is that in ERMHD
and FLR-MHD the sign-indefinite invariant is the one ``with fewer powers
of~$k_\perp$'' (i.e., ${\cal H}_{\rm mf}$), whereas in 3D~hydro the
sign-indefinite invariant is the one with ``more powers of~$k$'' (the helicity).

If (\ref{eq:AB1}) is not
satisfied for an invariant~$B$, then we will assume that~$B$ cascades to
larger~$k$. This assumption recovers the well known forward cascades of energy in 3D hydro, enstrophy in 2D~hydro, KAW
energy in ERMHD and FLR-MHD at $k_\perp \rho_i \gg 1$, and energy and
cross helicity in MHD and RMHD. In~\S\ref{sec:HBITEF}, we will apply
this assumption to determine the
cascade directions of ${\cal H}_{\textrm{ph-sp}}$ and~$W_{\rm compr}$ in~KRMHD 
and~${\cal H}_{\textrm{ph-sp}}$ and~$W_{h_i}$ in~ERMHD, and in
Appendix~\ref{ap:phasemixing} we will invoke this assumption when
discussing parallel phase mixing in KRMHD.

\subsection{The cascade of gyrokinetic helicity in the ITEF approximation}
\label{sec:HBITEF} 

At $k_\perp \rho_i \ll 1$, the ITEF approximation reduces to KRMHD
(\S\ref{sec:KRMHD}). In this regime, the free energy divides into
separately conserved Alfv\'enic and compressive
components ($W_{\rm AW}$ and~$W_{\rm compr}$, which are defined in (\ref{eq:KRMHDWAW})
and~(\ref{eq:KRMHDWcompr})), and the gyrokinetic helicity divides into
separately conserved magnetofluid and phase-space
components (${\cal H}_{\rm mf}$ and~${\cal H}_{\textrm{ph-sp}}$, which are
defined in~(\ref{eq:HAKRMHD}) and~(\ref{eq:HnAKRMHD})). The invariants
$W_{\rm AW}$ and~${\cal H}_{\rm mf}$ are integrals of the same
fluctuating quantities ($\varphi$ and~$A$) that contain the same
number of gradient operators and satisfy neither
(\ref{eq:AB1}) nor~(\ref{eq:AB2}). An equivalent statement holds for
$W_{\rm compr}$ and~${\cal H}_{\textrm{ph-sp}}$. If these four
quantities are injected at $k_\perp \rho_i \ll 1$, then all four
cascade to smaller scales, provided there is enough dissipation at
$k_\perp \rho_i < 1$ to truncate the cascades within the KRMHD regime
(thereby making the turbulence independent of cascade directions at
$k_\perp \rho_i > 1$).

At $k_\perp \rho_i \gg 1$, the ITEF approximation reduces to ERMHD
(\S\ref{sec:ERMHD}). In ERMHD, the free energy divides into separately
conserved KAW and ion-entropy components ($W_{\rm KAW}$ and~$W_{h_i}$,
which are defined in~(\ref{eq:WKAW}) and~(\ref{eq:Whi})), and the gyrokinetic helicity
again divides into separately conserved magnetofluid and phase-space
components (${\cal H}_{\rm mf}$ and~${\cal H}_{\textrm{ph-sp}}$, which
are defined in~(\ref{eq:HAERMHD}) and~(\ref{eq:HnAERMHD})). The
invariants~$W_{h_i}$ and~${\cal H}_{\textrm{ph-sp}}$ are integrals of
the same fluctuating quantity ($h_i$) that contain the same number of
gradient operators (zero) and satisfy neither~(\ref{eq:AB1})
nor~(\ref{eq:AB2}). We thus expect both to cascade to larger~$k_\perp$. 
The invariants $W_{\rm KAW}$ and ${\cal H}_{\rm mf}$ are
likewise integrals of the same fluctuating quantities ($\varphi$
and~$A$) and satisfy~(\ref{eq:AB1}),  with~B being $W_{\rm KAW}$
and~A being~${\cal H}_{\rm mf}$, as discussed
in~\S\ref{sec:Fjortoft}.  As a consequence, $W_{\rm KAW}$ 
cascades to smaller scales, and ${\cal H}_{\rm mf}$ undergoes an
inverse cascade \citep{schekochihin09}, just like~$W$
and~${\cal H}_{\rm mf}$ in FLR-MHD at
$k_\perp \rho_i \gg 1$~\citep{meyrand21}.

\begin{figure}
\centerline{
  \includegraphics[width=10cm]{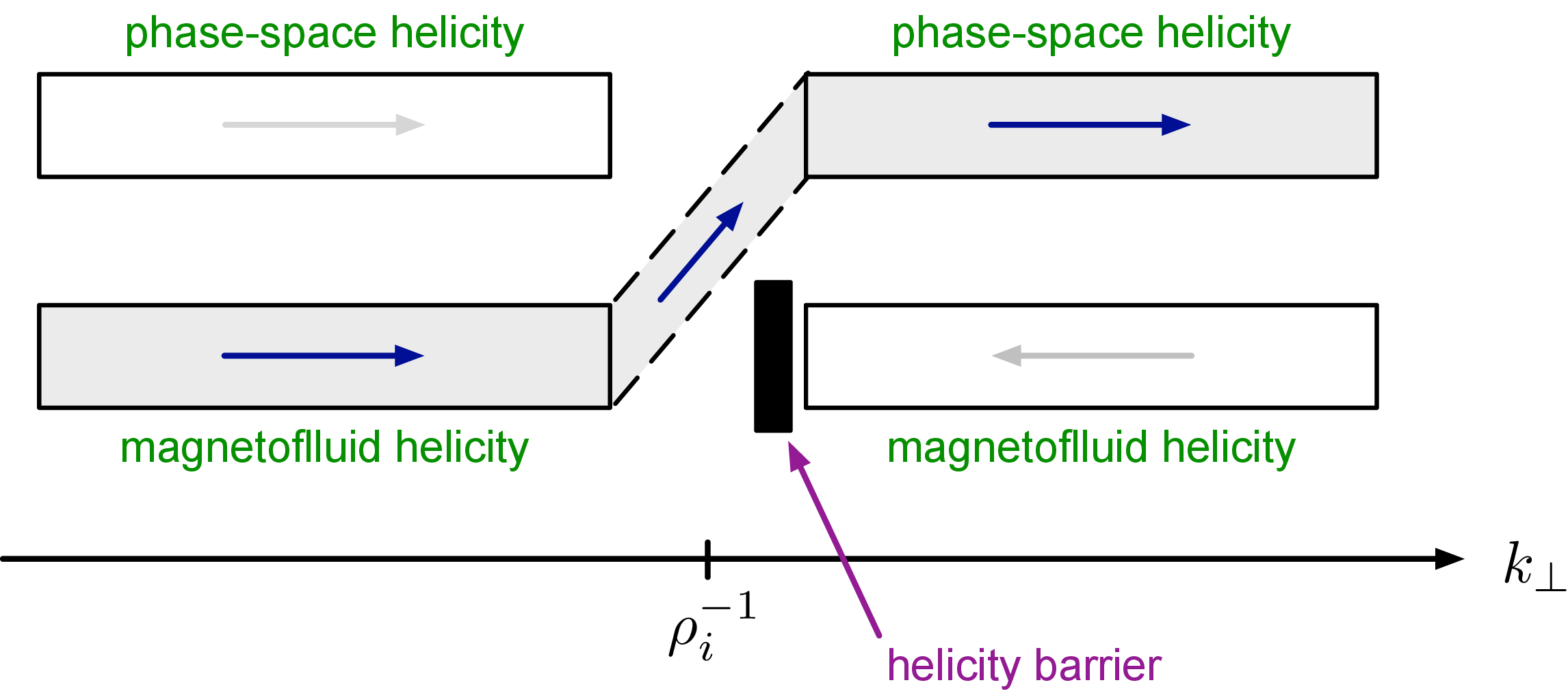}
  }
  \caption{If gyrokinetic helicity is injected as magnetofluid
    helicity at $k_\perp \rho_i \ll
    1$, it can cascade to~$k_\perp \rho_i \sim 1$,
    transform into phase-space helicity, and then continue cascading
    to~$k_\perp \rho_i \gg 1$. Horizontal arrows indicate the cascade
    directions of the two helicity types.}
  \label{fig:H_cascade} 
\end{figure}

At $k_\perp \rho_i \sim 1$, nonlinear interactions can convert one
type of free energy into another \citep{schekochihin09} and one type
of helicity into another.  Magnetofluid helicity~${\cal H}_{\rm mf}$
injected (initially or steadily) at $k_\perp \rho_i \ll 1$ can cascade
to $k_\perp \rho_i \sim 1$, transform into phase-space
helicity~${\cal H}_{\textrm{ph-sp}}$ at the rate given
by~(\ref{eq:dH1dt0}), and cascade to~$k_\perp \rho_i \gg 1$, as
illustrated in Figure~\ref{fig:H_cascade}. At $k_\perp \rho_i \gg 1$,
nonlinear perpendicular phase mixing \citep{schekochihin09}
transports~${\cal H}_{\textrm{ph-sp}}$ to small scales in velocity
space at which~${\cal H}_{\textrm{ph-sp}}$ dissipates via ion
collisions. The forward cascade of gyrokinetic helicity illustrated in
Figure~\ref{fig:H_cascade}  increases the
forward cascade rate of~$W_{\rm AW}$ and the turbulent heating rate in
imbalanced Alfv\'enic turbulence, in which one of the Elsasser
variables~$\bm{z}^\pm = \bm{u}_\perp \pm \delta
\bm{B}_\perp/\sqrt{4\pi n_i m_i}$ is much larger than the other,
because it enables the energy cascade rate of the dominant Elsasser
variable to exceed the energy cascade rate of the sub-dominant
Elsasser variable.

When Alfv\'enic free energy~$W_{\rm AW}$, magnetofluid
helicity~${\cal H}_{\rm mf}$, and compressive free
energy~$W_{\rm compr}$ are injected at $k_\perp \rho_i \ll 1$, the
cascade of~$W_{\rm compr}$ from $k_\perp \rho_i \ll 1$
to~$k_\perp \rho_i \sim 1$ enhances the turbulent heating rate in two
ways. First, it directly feeds the cascade of~$W_{h_i}$ (the ion
entropy cascade) at $k_\perp \rho_i \gg 1$, which dissipates via ion
collisions and leads to ion heating \citep{schekochihin09}. Second, it
increases the amplitudes of the compressive fluctuations $\delta n/n$,
$\delta B/B$, and~$g_i$ at $k_\perp \rho_i \sim 1$, which increases
$\mathrm{d}{\cal H}_{\rm mf}/\mathrm{d} t$ in (\ref{eq:dH1dt0}).  This
in turn enables more magnetofluid helicity~${\cal H}_{\rm mf}$ at
$k_\perp \rho_i \ll 1$ to cascade to $k_\perp \rho_i \sim 1$, which
increases the energy cascade rate of the dominant Elsasser variable.
This two-fold enhancement of turbulent heating by $W_{\rm compr}$
injection likely plays an important role in the near-Sun solar wind
and coronal holes (regions of the solar corona that are magnetically
connected to the solar surface at one end and the solar wind at the
other) as these regions are permeated by imbalanced Alfv\'en-wave
turbulence \citep{dmitruk02,cranmer05,depontieu07,verdini07} and
density fluctuations~\citep{coles89,harmon05,cuesta23} that can have large
amplitudes near the coronal base~\citep{raymond14}.

\section{Conclusion}
\label{sec:conclusion}

Our principal finding is that astrophysical gyrokinetics possesses a second
three-dimensional quadratic invariant in addition to the free energy. We call this new
invariant the gyrokinetic helicity~${\cal H}$ and obtain the limiting
forms of~$\mathrm{Re}\,{\cal H}$ in several subsidiary regimes of astrophysical
gyrokinetics. We find it useful to subdivide~$\mathrm{Re}\,{\cal H}$
into two components: the magnetofluid helicity~${\cal H}_{\rm mf}$ and the
phase-space helicity~${\cal H}_{\textrm{ph-sp}}$. The magnetofluid
helicity results from the magnetic fluctuations and low-order moments
of the ion and electron distribution functions. This part of the
helicity is equivalent to the RMHD cross helicity in kinetic reduced
magnetohydrodynamics (KRMHD), the magnetic helicity in electron
magnetohydrodynamics (ERMHD), and the generalized helicity in both
finite-Larmor-radius magnetohydrodynamics (FLR-MHD) and the kinetic
reduced electron heating model~(KREHM). The phase-space helicity
depends upon the small-scale velocity-space structure of the ion and
electron distribution functions.  In KRMHD and ERMHD,
${\cal H}_{\rm mf}$ and~${\cal H}_{\textrm{ph-sp}}$ are separately
conserved. 

Our findings provide a framework for understanding how the helicity
barrier of FLR-MHD~\citep{meyrand21} is modified by non-Alfv\'enic
fluctuations.  In FLR-MHD, ${\cal H}_{\textrm{ph-sp}}=0$ by
assumption, and ${\cal H}_{\rm mf}$ can cascade only towards
larger~$k_\perp$ at~$k_\perp \rho_i \ll 1$ and only towards
smaller~$k_\perp$ at~$k_\perp \rho_i \gg 1$. As a consequence,
if~${\cal H}_{\rm mf}$ is steadily injected at~$k_\perp \rho_i \ll 1$
and dissipates only at~$k_\perp \rho_i \gg 1$,
then~${\cal H}_{\rm mf}$ cannot cascade from the injection scale to
the dissipation scale and hence builds up secularly in
time~\citep{meyrand21,squire22}.  In the more general isothermal
electron fluid (ITEF) approximation, when~${\cal H}$ is injected as
magnetofluid helicity at~$k_\perp \rho_i \ll 1$, it can cascade
to~$k_\perp \rho_i \sim 1$, transform into phase-space helicity at
$k_\perp \rho_i \sim 1$, and cascade to~$k_\perp \rho_i \gg 1$, as
illustrated in figure~\ref{fig:H_cascade}. At $k_\perp \rho_i \gg 1$,
perpendicular phase mixing~\citep{schekochihin09} transfers~${\cal
  H}_{\textrm{ph-sp}}$ to small scales in velocity space, and~${\cal
  H}_{\textrm{ph-sp}}$ dissipates via ion collisions. This forward
cascade of gyrokinetic helicity
bypasses the helicity barrier, increases the
turbulent heating rate over the FLR-MHD value, and may play an important role in coronal
holes and the near-Sun solar wind.

Our results suggest some potentially fruitful directions for future
research, including exploring the implications of
$\mathrm{Im}{\cal H}$ conservation, using (\ref{eq:dH1dt0}) to
investigate the possible breakdown of the helicity barrier as $\beta$
increases to values~$\gtrsim 1$, and investigating whether~${\cal H}$
or some related quantity is a quadratic invariant in inhomogeneous
plasmas with more complicated magnetic geometries, such as tokamak and
stellarator plasmas.  Another question raised by our results is
whether gyrokinetic helicity conservation leads to a cascade
bottleneck in velocity space. The discussion at the end of
Appendix~\ref{ap:phasemixing} begins to address this question 
for KRMHD, but a more comprehensive investigation is needed.
Finally, it would be useful to evaluate the degree to
which the conversion of ${\cal H}_{\rm mf}$
into~${\cal H}_{\textrm{ph-sp}}$ at $k_\perp \rho_i\sim 1$ enhances
the turbulent heating rate in imbalanced turbulence. We have taken a
step in this direction via~(\ref{eq:dH1dt0}), which gives an analytic
expression for~$\mathrm{d}{\cal H}_{\rm mf}/\mathrm{d}t$. Quantifying
how the resulting heating-rate modification depends upon the
properties of the turbulence at the outer scale (injection scale)
would contribute significantly to the development of more rigorous,
physics-based models of coronal-hole heating and solar-wind
generation.

\vspace{0.2cm} 
\noindent {\bf Acknowledgements.} We thank Toby Adkins, Alexander
Schekochihin, and Jonathan Squire for valuable discussions.

\vspace{0.2cm} 
\noindent {\bf Funding.} This work was supported by the
U.S. Department of Energy (B.C., grant number DE-SC0026201; R.M., grant number DE-SC0026201); and the
National Aeronautics and Space Administration (B.C., grant numbers
80NSSC24K0171 and NNN06AA01C; A.M., grant number NNN06AA01C; R.M. grant number
80NSSC24K0171).

\vspace{0.2cm} 
\noindent {\bf Authors' statement on artificial intelligence.} No artificial intelligence
was used to obtain or check these results or write or edit this paper.

\vspace{0.2cm} 
\noindent {\bf Declaration of interests.} The authors report no
conflict of interest.

\vspace{0.2cm} 
\noindent {\bf Author ORCID.}
B.~Chandran, 0000-0003-4177-3328;
A.~Mallet, 0000-0001-9202-1340;
R.~Meyrand, 0000-0002-8327-5848

\appendix

\section{Calculation of~$\mathrm{d}{\cal H}_{\rm mf}/\mathrm{d}t$
  and~$\mathrm{d}{\cal H}_{\textrm{\rm ph-sp} }/\mathrm{d}t$ in the isothermal
  electron fluid approximation}
\label{ap:dHdt}

As shown by \cite{schekochihin19}, integrating~(\ref{eq:dgdt}) over velocity at
fixed~$\bm{r}$, dividing by~$n$, and subtracting~(\ref{eq:econt}) 
yields
\[
  \frac{\partial }{\partial t} \left[
    \left(1 - \hat{\Gamma}_0 \right)\varphi + \left(1 - \hat{\Gamma}_1   \right) \frac{\delta B}{B}
 \right]
 =
 -\frac{\rho_{i } v_{\rm th \it i}}{2} \left\{ \varphi,  \left(1 - \hat{\Gamma}_0 \right)\varphi + \left(1 - \hat{\Gamma}_1   \right) \frac{\delta B}{B}   \right\}
\]
\[
  +\frac{v_{\rm th \it i}}{\beta_{i }} \nabla_\parallel \hat{\nabla}_\perp^2 A
  + \rho_{i }\overline{ \left\langle \left\{ \langle A \rangle_{\bm{R}_i} - A, v_\parallel g_i   \right\}\right\rangle_{\bm{r}} }
\]
\begin{equation}
  - \frac{\rho_{i } v_{\rm th \it i}}{2} \left[
\overline{
\left\langle 
  \left\{\langle \varphi \rangle_{\bm{R}_i} - \varphi, g_i - \hat{v}_\perp^2 \hat{J}_1 \frac{\delta B}{B} F_{0i}\right\} +
  \left\{ \hat{v}_\perp^2 \hat{J}_1 \frac{\delta B}{B}, g_i\right\} 
\right\rangle_{\bm{r}}}
- \frac{Z}{\tau}\left\{\frac{\delta n}{n}, \frac{\delta B}{B}  \right\}
    \right].
\label{eq:dgdt_zero} 
\end{equation}
To obtain~(\ref{eq:dH1dt0}), we take the time derivative
of~(\ref{eq:H1}) and evaluate partial time derivatives using
(\ref{eq:dAdt}) and~(\ref{eq:dgdt_zero}).  Defining
$N = \left(1 - \hat{\Gamma}_0\right) \varphi + \left(1 -
  \hat{\Gamma}_1\right) \displaystyle\frac{\delta B}{B}$, we write the
resulting equation in the form
\[
\frac{1}{m_in_i v_{\rm th \it i}} \frac{\mathrm{d}}{\mathrm{d}t}{\cal
  H}_{\rm mf} =\int \frac{\mathrm{d}^3 \bm{r}}{V} \Biggr(
  - \stackrel{\circled{1}}{ \frac{v_{\rm th \it i}}{2} \left[\left(1 - \hat{\Gamma}_0\right) \varphi \right] \frac{\partial \varphi}{\partial z}}
  - \stackrel{\circled{2}}{  \frac{v_{\rm th \it i}}{2} \left[\left(1
        - \hat{\Gamma}_1\right) \frac{\delta B}{B}
    \right]\frac{\partial \varphi}{\partial z}}
\]
\[
  + \stackrel{\circled{3}}{ \frac{\rho_{i }v_{\rm th \it i}}{2} N \{A, \varphi\}}
 \;+\; \frac{v_{\rm th \it i}}{2} \frac{Z}{\tau} \Biggr[ \stackrel{\raisebox{5pt}{\circled{4}}}{\left(1 - \hat{\Gamma}_0\right)\varphi}
    + \stackrel{\raisebox{7pt}{\circled{5}}}{\frac{\delta B}{B}}
    - \stackrel{\raisebox{7pt}{\circled{6}}}{\hat{\Gamma}_1\frac{\delta B}{B}}
    \Biggr] \frac{\partial }{\partial z} \frac{\delta n}{n}
  \]
  \[
\; -\;  \frac{\rho_{i }v_{\rm th \it i}}{2} \frac{Z}{\tau}  \Biggr[ \stackrel{\raisebox{5pt}{\circled{7}}}{\left(1 - \hat{\Gamma}_0\right)\varphi}
    + \stackrel{\raisebox{7pt}{\circled{8}}}{\frac{\delta B}{B}}
    - \stackrel{\raisebox{7pt}{\circled{9}}}{\hat{\Gamma}_1\frac{\delta B}{B}}
  \Biggr] \left\{ A, \frac{\delta n}{n}\right\}   \; - \;\stackrel{\circled{10}}{  \frac{\rho_{i }v_{\rm th \it i}}{2} A \left\{ \varphi, N\right\}}
\]
\[
  + \stackrel{\circled{11}}{  \frac{v_{\rm th \it i}}{\beta_{i }} A \nabla_\parallel \hat{\nabla}_\perp^2 A}
  \;+\; \stackrel{\raisebox{7pt}{\circled{12}}}{  \rho_{i } A \overline{\left\langle\left\{\langle A \rangle_{\bm{R}_i} - A, v_\parallel g_i\right\}  \right\rangle_{\bm{r}}}}
  \;+\;\stackrel{\raisebox{7pt}{\circled{13}}}{\frac{\rho_{i } v_{\rm th \it i}}{2} \frac{Z}{\tau} A \left\{\frac{\delta n}{n}, \frac{\delta B}{B} \right\}  }
\]
\begin{equation} 
  - \stackrel{\raisebox{7pt}{\circled{14}}}{\frac{\rho_{i }v_{\rm th \it i}}{2} A \overline{\left\langle \left\{ \langle\varphi \rangle_{\bm{R}_i} - \varphi, g_i - \hat{v}_\perp^2 \hat{J}_1 \frac{\delta B}{B} F_{0i}  \right\}\right\rangle_{\bm{r} }}}
    \;  -\; \stackrel{\raisebox{7pt}{\circled{15}}}{\frac{\rho_{i }v_{\rm th \it i}}{2} A \overline{\left\langle \left\{ \hat{v}_\perp^2 \hat{J}_1 \frac{\delta B}{B},g_i  \right\}\right\rangle_{\bm{r} }}}
\Biggr).
\label{eq:dH1dta} 
\end{equation} 

Three of the terms on the right-hand side of (\ref{eq:dH1dta}) vanish
individually. Term~$\circled{1}$ vanishes because the operator
$1 - \hat{\Gamma}_0$ is an infinite series in powers
of~$\nabla_\perp^2$, and the integrand can be rewritten as a total
derivative after half the~$\grad_\perp$ operators are integrated by
parts.  Integrating the $\partial /\partial z$ term in~$\circled{11}$
by parts, we find that
\begin{align}
  \circled{11} & = \frac{v_{\rm th \it i}}{\beta_{i }}
                 \int\frac{\mathrm{d}^3 \bm{r}}{V} \left[ -
                 \left(\hat{\grad}_\perp A\right) \cdot
                 \frac{\partial }{\partial z} \hat{\grad}_\perp A - A \rho_i
                 \{ A, \hat{\nabla}_\perp^2 A\}\right] \label{eq:11a} 
\\
  & = -\frac{1}{2}\frac{v_{\rm th \it i}}{\beta_{i }} \int\frac{\mathrm{d}^3 \bm{r}}{V}
     \frac{\partial }{\partial z}
    \left|\hat{\grad}_\perp A\right|^2  = 0\label{eq:11b}, 
\end{align}
where we made use of the identity
\begin{equation}
  \int \frac{\mathrm{d}^3 \bm{r}}{V} f \{f, g\} = \int \frac{\mathrm{d}^3 \bm{r}}{V} \left(\bm{\hat{z}} \times \grad_\perp\frac{f^2}{2}\right) \cdot \grad_\perp g 
  = \int \frac{\mathrm{d}^3 \bm{r}}{V} \grad_\perp \cdot \left[ \left(\bm{\hat{z}} \times \grad_\perp \frac{f^2}{2}\right)g\right] = 0
    \label{eq:purediv}
  \end{equation}
(as~$V\rightarrow\infty$) to drop the second term
in the integrand in~(\ref{eq:11a}).
Switching the order of integration and changing integration variables from~$(\bm{r},\bm{v})$
to~$(\bm{R}_i, \bm{v})$, we obtain
\begin{align}
\circled{12} &= \rho_{i }\int\frac{\mathrm{d}^3\bm{v}}{n_i} \int\frac{\mathrm{d}^3
               \bm{r}}{V} 
\left(A \left\langle \left\{
               \langle{A}\rangle_{\bm{R}_i},v_\parallel g_i
               \right\}\right\rangle_{\bm{r}}  
               -
A \left\{A, v_\parallel\left\langle g \right\rangle_{\bm{r}}  
               \right\}
               \right) \label{eq:12a} \\
  & =  \rho_{i } \int\frac{\mathrm{d}^3\bm{v}}{n_i} \int\frac{\mathrm{d}^3
    \bm{R}_i}{V} \langle A\rangle_{\bm{R}_i} \left\{\langle A
    \rangle_{\bm{R}_i},v_\parallel g_i  \right\} = 0, \label{eq:12b} 
\end{align}
where we invoked~(\ref{eq:purediv})~twice.

Four additional terms on the right-hand side of (\ref{eq:dH1dta}) vanish
in groups of two.
Term~$\circled{3}$ cancels term~$\circled{10}$ because
\begin{equation}
  \int \frac{\mathrm{d}^3 \bm{r}}{V} f \{g, h\} = \int\frac{\mathrm{d}^3 \bm{r}}{V} f \grad_\perp \cdot (h \bm{\hat{z}} \times \grad_\perp g) =
  - \int\frac{\mathrm{d}^3 \bm{r}}{V} h (\bm{\hat{z}} \times \grad_\perp g)\cdot \grad_\perp f
  =  \int\frac{\mathrm{d}^3 \bm{r}}{V} h \{f, g\}.
  \label{eq:cyclic}
\end{equation}
That is, arbitrary functions $f$, $g$, and~$h$ in an integrand of the
form $f \{ g,h\}$ within a volume integral can be cyclically
permuted. Terms~$\circled{8}$ and~$\circled{13}$ cancel for the same
reason.

The remaining terms do not vanish but can be combined into a more
compact form. Terms $\circled{4}$, $\circled{6}$, $\circled{7}$, and $\circled{9}$
(re)combine straightforwardly:
\begin{equation}
  \circled{4} + \circled{6} + \circled{7} + \circled{9} = \frac{v_{\rm
      th\it i} Z}{2 \tau} \int \frac{\mathrm{d}^3\bm{r}}{V} \left[\left( 1
      - \hat{\Gamma}_0\right)\varphi - \hat{\Gamma}_1
    \frac{\delta B}{B} \right] \nabla_\parallel \frac{\delta
    n}{n}.
  \label{eq:4769}
\end{equation}
As $1 - \hat{\Gamma}_1$ is a series in powers of $\nabla_\perp^2$, we
can integrate by parts the $\grad_\perp$ operators in
term~$\circled{2}$ an even number of times to obtain
\begin{equation}
  \circled{2}   = - \frac{v_{\rm th \it i}}{2} \int \frac{\mathrm{d}^3 \bm{r}}{V} \frac{\delta B}{B} \frac{\partial }{\partial z} \left[\left(1 - \hat{\Gamma}_1 \right) \varphi \right] .
  \label{eq:term2}
\end{equation}
With the use of (\ref{eq:dB}), we then find that
\begin{align}
  \circled{2} + \circled{5} &= - \frac{v_{\rm th \it i}}{2} \int \frac{\mathrm{d}^3 \bm{r}}{V} \frac{ \delta B}{B} \frac{\partial}{\partial z}
  \left[ \frac{2}{\beta_{i }} \frac{\delta B}{B} +
                              \overline{\hat{v}_\perp^2 \hat{J}_1
                              g_i(\bm{r},v_\perp,v_\parallel,t)}\right] \label{eq:2plus5a}
  \\
  & =  - \frac{v_{\rm th \it i}}{2}  \int\frac{\mathrm{d}^3\bm{v}}{n_i} \int \frac{\mathrm{d}^3 \bm{r}}{V}  \hat{v}_\perp^2\left[ \hat{J}_1\frac{ \delta B}{B}\right] \frac{\partial }{\partial z}g_i(\bm{r},v_\perp,v_\parallel,t),
\label{eq:2plus5c}  \\
  & =  - \frac{v_{\rm th \it i}}{2}  \int\frac{\mathrm{d}^3\bm{v}}{n_i} \int \frac{\mathrm{d}^3 \bm{R}_i}{V}  \hat{v}_\perp^2\left[ \hat{J}_1\frac{ \delta B(\bm{R}_i,t)}{B}\right] \frac{\partial }{\partial z}g_i(\bm{R}_i,v_\perp,v_\parallel,t),
\label{eq:2plus5d} 
\end{align}
where we integrated by parts to go from
(\ref{eq:2plus5a}) to~(\ref{eq:2plus5c}) and made use of the fact that
$\bm{r}$ in~(\ref{eq:2plus5c}) is a dummy variable that we can
re-name~$\bm{R}_i$ to obtain~(\ref{eq:2plus5d}). We now
regard~$\bm{R}_i$ in~(\ref{eq:2plus5d}) as the ion guiding center and change
variables from~$\bm{R}_i$ to~$\bm{r}$ to obtain
\begin{align}
  \circled{2} + \circled{5} &
=  - \frac{v_{\rm th \it i}}{2}  \int\frac{\mathrm{d}^3\bm{v}}{n_i} \int
                              \frac{\mathrm{d}^3 \bm{r}}{V}
                              \Biggr\langle
 \hat{v}_\perp^2\left[ \hat{J}_1\frac{ \delta B(\bm{R}_i,t)}{B}\right] \frac{\partial }{\partial z}g_i(\bm{R}_i,v_\perp,v_\parallel,t)
                              \Biggr\rangle_{\bm{r}}.
\label{eq:2plus5e} 
\end{align}
Equation~(\ref{eq:2plus5e})  can alternatively be obtained
from~(\ref{eq:2plus5c}) by noting that the integrand
of~(\ref{eq:2plus5e})  is $\hat{J}_0(a)$ operating upon the entire integrand
of~(\ref{eq:2plus5c}) and that when one expands that $\hat{J}_0(a)$ in powers
of~$a$ all nonzero powers integrate to zero.
Switching the order of integration in term~$\circled{15}$, we obtain
\begin{align}
  \circled{15} & = - \frac{\rho_i v_{\rm th \it i}}{2} \int
                 \frac{\mathrm{d}^3 \bm{v}}{n_i} \Biggr\langle \int
                 \frac{\mathrm{d}^3 \bm{r}}{V} A(\bm{r},t) \left\{
                 \hat{v}_\perp^2 \hat{J}_1 \frac{\delta B(\bm{R}_i,t)}{B},
                 g_i(\bm{R}_i, v_\perp,
                 v_\parallel,t)\right\}\Biggr\rangle_{\bm{r}} \label{eq:15a}\\
  & = \frac{\rho_i v_{\rm th \it i}}{2} \int\frac{\mathrm{d}^3\bm{v}}{n_i}
    \int\frac{\mathrm{d}^3 \bm{r}}{V} \Biggr\langle
    \left[\hat{v}_\perp^2 \hat{J}_1 \frac{\delta B(\bm{R}_i,t)}{B}\right] \left\{
    A(\bm{r},t),g_i(\bm{R}_i,v_\perp,v_\parallel,t)\right\}\Biggr\rangle_{\bm{r}},
    \label{eq:15b} 
\end{align}
where we have made use of~(\ref{eq:cyclic}) to go from~(\ref{eq:15a})
to~(\ref{eq:15b}). Adding (\ref{eq:2plus5e}) and~(\ref{eq:15b}), we
find that
\begin{equation}
  \circled{2} + \circled{5} + \circled{15} = - \frac{v_{\rm th \it i}}{2}
  \int\frac{\mathrm{d}^3\bm{v}}{n_i} \int\frac{\mathrm{d}^3\bm{r}}{V}
  \Biggr\langle
\left[\hat{v}_\perp^2 \hat{J}_1 \frac{\delta B(\bm{R}_i,t)}{B}\right]
\nabla_\parallel g_i(\bm{R}_i,v_\perp,v_\parallel,t)
  \Biggr\rangle_{\bm{r}},
  \label{eq:2plus5plus15}
\end{equation}
where the~$A$ appearing in the~$\nabla_\parallel$
from~(\ref{eq:gradpar}) is evaluated at~$\bm{r}$.

The final term in~$\mathrm{d}{\cal H}_{\rm mf}/\mathrm{d}t$ is
term~$\circled{14}$, which
we rewrite as
\begin{align}
&  \circled{14}  =  - \frac{\rho_i v_{\rm th \it i}}{2}
                 \int\frac{\mathrm{d}^3 \bm{v}}{n_i} \Biggr\langle \int
                 \frac{\mathrm{d}^3\bm{r}}{V}
 \left(g_i - \hat{v}_\perp^2
                 \hat{J}_1 \frac{\delta B}{B} F_{0i}  \right)
                 \Big\{ \varphi(\bm{r}) - \langle
                 \varphi\rangle_{\bm{R}_i}, A(\bm{r})\Big\} \label{eq:14a} 
  \Biggr\rangle_{\bm{r}}\\
               & =  - \frac{\rho_i v_{\rm th \it i}}{2}
                 \int\frac{\mathrm{d}^3 \bm{v}}{n_i} \int
                 \frac{\mathrm{d}^3\bm{R}_i}{V}
  \Biggr\langle   \left(g_i - \hat{v}_\perp^2
                 \hat{J}_1 \frac{\delta B}{B} F_{0i}  \right) \Big\{ \varphi(\bm{r}) - \langle
                 \varphi\rangle_{\bm{R}_i}, A(\bm{r})\Big\}
                 \Biggr\rangle_{\bm{R}_i} \label{eq:14b} \\
               & =  - \frac{\rho_i v_{\rm th \it i}}{2}
                 \int\frac{\mathrm{d}^3 \bm{v}}{n_i} \int
                 \frac{\mathrm{d}^3\bm{R}_i}{V}
  \Biggr\langle  \left(g_i - \hat{v}_\perp^2
                 \hat{J}_1 \frac{\delta B}{B} F_{0i}  \right)
                 \Big[ \big\{ \varphi({\bm{r}}),A(\bm{r}) \big\}  -
                 \big\{\langle \varphi\rangle_{\bm{R}_i},\langle
                 A\rangle_{\bm{R}_i} \big\} \Big] \Biggr\rangle_{\bm{R}_i}
  \label{eq:14c} \\
  & =  - \frac{\rho_i v_{\rm th \it i}}{2} \int\frac{\mathrm{d}^3 \bm{v}}{n_i} \int \frac{\mathrm{d}^3\bm{r}}{V}
\Biggr\langle   \left(g_i - \hat{v}_\perp^2
                 \hat{J}_1 \frac{\delta B}{B} F_{0i}  \right)
                 \Big[ \big\{ \varphi({\bm{r}}),A(\bm{r}) \big\}  -
                 \big\{\langle \varphi\rangle_{\bm{R}_i},\langle
    A\rangle_{\bm{R}_i} \big\} \Big] \Biggr\rangle_{\bm{r}} \label{eq:14d} 
\end{align}
where we have interchanged the order of integration and
invoked~(\ref{eq:cyclic}) to obtain~(\ref{eq:14a}), changed
integration variables from~$\bm{r}$ to~$\bm{R}_i$ to go
from~(\ref{eq:14a}) to~(\ref{eq:14b}), and then changed back
to go from~(\ref{eq:14c})  to~(\ref{eq:14d}). The
arguments of each~$g_i$ in (\ref{eq:14a}) through~(\ref{eq:14d})
are~$(\bm{R}_i,v_\perp,v_\parallel,t)$,
the arguments of each~$\delta B$ are~$(\bm{R}_i,t)$, and we have for brevity
omitted~$t$ when writing the functional arguments of~$\varphi$ and~$A$.
Equations~(\ref{eq:11b}) through (\ref{eq:14d}) enable us to
rewrite~(\ref{eq:dH1dta}) as~(\ref{eq:dH1dt0}).
    
Upon taking the time derivative of~(\ref{eq:H2}), making use
of~(\ref{eq:dgdt}), and invoking~(\ref{eq:purediv})  to
drop terms of the form $V^{-1}\int \mathrm{d}^3\bm{R}_i f\{f,g\}$, we obtain 
\[
  \frac{\mathrm{d}{\cal H}_{\rm \textrm{ph-sp} }}{\mathrm{d}t} =
  n_i T_i \int\frac{\mathrm{d}^3\bm{R}_i}{V} \int\frac{\mathrm{d}^3\bm{v}}{n_i}
\Biggr[
  - \stackrel{\circled{16}}{\frac{Z}{\tau}g_i \frac{\partial }{\partial z} \left\langle \frac{\delta n}{n} \right\rangle_{\bm{R}_i}}
  + \stackrel{\circled{17}}{\hat{v}_\perp^2 \left( \hat{J}_1 \frac{\delta B}{B}\right) \frac{\partial}{\partial z} g_i}
 \]
\[
 + \stackrel{\circled{18}}{\frac{Z}{\tau} F_{0i} \left(\hat{J}_1 \hat{v}_\perp^2 \frac{\delta B}{B}\right) \frac{\partial}{\partial z}
  \left\langle \frac{\delta n}{n}\right\rangle_{\bm{R}_i}}
  +\stackrel{\circled{19}}{ \frac{Z}{\tau} \rho_{i } g_i \left\langle\left\{A, \frac{\delta n}{n}  \right\} \right\rangle_{\bm{R}_i} }
  - \stackrel{\circled{20}}{ \rho_{i } \hat{v}_\perp^2 \left(\hat{J}_1 \frac{\delta B}{B} \right)
   \left\langle\left\{A, g_i  \right\} \right\rangle_{\bm{R}_i}}
\]
\begin{equation}
- \stackrel{\circled{21}}{ \frac{Z}{\tau} \rho_{i } F_{0i} \left(\hat{v}_\perp^2 \hat{J}_1 \frac{\delta B}{B} \right) \left\langle\left\{A, \frac{\delta n}{n}  \right\} \right\rangle_{\bm{R}_i} }
  - \stackrel{\circled{22}}{\rho_{i } \left( g_i - \hat{v}_\perp^2 \hat{J}_1 \frac{\delta B}{B} F_{0i}\right) \left\langle\left\{
        A - \langle A\rangle_{\bm{R}_i}, \varphi - \langle \varphi \rangle_{\bm{R}_i}\right\} \right\rangle_{\bm{R}_i} } \Biggr].
  \label{eq:dH2dta} 
\end{equation}
Following a path similar to the one from~(\ref{eq:dH1dta})
to~(\ref{eq:dH1dt0}), we rewrite~(\ref{eq:dH2dta}) as
\begin{align}
  \frac{\mathrm{d}}{\mathrm{d}t} & {\cal H}_{\rm \textrm{ph-sp} }  = \,  n_i T_i
                                             \int\frac{\mathrm{d}^3\bm{r}}{V}\frac{Z}{\tau}
  \left(\hat{\Gamma}_1 \frac{\delta B}{B}\right) \nabla_\parallel
                                             \frac{\delta n}{n}\nonumber \\
  &  + n_i T_i
    \int\frac{\mathrm{d}^3\bm{r}}{V} \int\frac{\mathrm{d}^3 \bm{v}}{n_i}
    \Biggr\langle \left[\frac{Z}{\tau} \frac{\delta n(\bm{r},t)}{n} +
    \hat{v}_\perp^2 \hat{J}_1 \frac{\delta B(\bm{R}_i,t)}{B}
    \right]\nabla_\parallel g_i(\bm{R}_i,v_\perp,v_\parallel,t)\nonumber \\
  & + \rho_i \left[g_i(\bm{R}_i,v_\perp,v_\parallel,t) - \hat{v}_\perp^2
    \hat{J}_1 \frac{\delta B(\bm{R}_i,t)}{B} F_{0i}   \right]
    \Big[ \big\{ \varphi({\bm{r}},t),A(\bm{r},t) \big\}  -
                 \big\{\langle \varphi\rangle_{\bm{R}_i},\langle
    A\rangle_{\bm{R}_i} \big\} \Big]
    \Biggr\rangle_{\bm{r}},
 \label{eq:dH2dtb}
\end{align}
where the $\left(\hat{\Gamma}_1 \delta B/B\right)\nabla_\parallel(\delta n/n)$ term on the right-hand side of~(\ref{eq:dH2dtb}) is the sum of
terms~$\circled{18}$ and~$\circled{21}$ in~(\ref{eq:dH2dta}),
the $(\delta n/n)\nabla_\parallel g$ term is the sum of
terms~$\circled{16}$ and~$\circled{19}$, the $(\hat{J}_1\delta
B/B)\nabla_\parallel g$ term
is the sum of terms~$\circled{17}$
and~$\circled{20}$,
the~$A$ appearing in $\nabla_\parallel g_i(\bm{R}_i, v_\perp,
v_\parallel, t)$ via (\ref{eq:gradpar})
is evaluated at~$\bm{r}$, and the bottom line of~(\ref{eq:dH2dtb}) is
another way of writing term~$\circled{22}$. Equation~(\ref{eq:dH2dt0}) 
follows from~(\ref{eq:dH1dt0})  and~(\ref{eq:dH2dtb}).

At~$k_\perp \rho_i \ll 1$, $\hat{\Gamma}_1 \rightarrow 1$, $\hat{J}_1
\rightarrow 1$, gyroaverages can be ignored, $\bm{R}_i \rightarrow
\bm{r}$, and (\ref{eq:dH2dtb}) becomes
\begin{equation}
  \frac{\mathrm{d}{\cal H}_{\rm \textrm{ph-sp} }}{\mathrm{d}t} =  n_i T_i
  \int \frac{\mathrm{d}^3\bm{r}}{V} \left( \frac{Z}{\tau} \frac{\delta
      B}{B} \nabla_\parallel \frac{\delta n}{n} + \frac{Z}{\tau}
    \frac{\delta n}{n} \nabla_\parallel \overline{ g_i} + \frac{\delta
      B}{B} \nabla_\parallel \overline{ \hat{v}_\perp^2 g_i}\right)
  \label{eq:dHentrdtc}
\end{equation}
to leading order in~$k_\perp \rho_i$, with corrections~$\sim
\mathrm{O}(k_\perp^2 \rho_i^2)$.
In this same large-wavelength limit, (\ref{eq:dn}) becomes $\delta n/n
= \overline{ g_i}$, and (\ref{eq:dB})  becomes $\overline{
  \hat{v}_\perp^2 g_i} =  - (Z/\tau) \delta n / n - (2/\beta_i)\delta
B/B$. The second term inside the parentheses on the right-hand side
of~(\ref{eq:dHentrdtc}) thus becomes $\propto \nabla_\parallel (\delta
n/n)^2$ and vanishes upon integration. The third term becomes the sum
of a term $\propto \nabla_\parallel (\delta B/B)^2$, which integrates
to~zero, and a term $\propto (\delta B/B) \nabla_\parallel (\delta
n/n)$, which cancels the first term inside the parentheses on the
right-hand side of~(\ref{eq:dHentrdtc}). Thus, $\mathrm{d}{\cal
  H}_{\textrm{ph-sp}}/\mathrm{d}t$ vanishes to leading order in
$k_\perp \rho_i$ at $k_\perp \rho_i \ll 1$,
as does~$\mathrm{d}{\cal
    H}_{\rm mf}/\mathrm{d}t = -\mathrm{d}{\cal
  H}_{\textrm{ph-sp}}/\mathrm{d}t$,
consistent with the
separate conservation of~${\cal H}_{\rm mf}$ and~${\cal
  H}_{\textrm{ph-sp}}$ in~KRMHD.
At $k_\perp \rho_i \gg 1$,
gyroaveraging (including the gyroaveraging
that went into the~$\hat{\Gamma}_1$ operator) reduces each term on the
right-hand side of~(\ref{eq:dH2dtb}) by a factor of~$\sim (k_\perp \rho_i)^{-1/2}$,
consistent with the separate conservation of~${\cal H}_{\rm mf}$ and~${\cal
  H}_{\textrm{ph-sp}}$ in~ERMHD.

\section{Additional quadratic invariants in KRMHD}
\label{ap:phasemixing} 

\cite{schekochihin09} showed that the compressive component of the
KRMHD free energy, $W_{\rm compr}$, can be further
subdivided into three energies that are separately conserved in the
absence of collisions and boundary effects. To show this,
they combined (\ref{eq:dnKRMHD}) and~(\ref{eq:perpAmpereKRMHD}) 
to obtain
\begin{equation}
  \frac{\delta n}{n} = \int_{-\infty}^\infty \mathrm{d}v_\parallel
  G_n,\qquad \frac{\delta B}{B} = \int_{-\infty}^\infty
  \mathrm{d}v_\parallel G_B,
  \label{eq:KRMHDdndB}
\end{equation}
where
\begin{equation} 
  G_n(\bm{r},v_\parallel,t) \;=\;  - \left[\frac{Z}{\tau} + 2\left(1 +
      \beta_{i }^{-1}\right)\right]^{-1} \frac{2\pi}{n_i}
  \int_0^\infty \mathrm{d}v_\perp\, v_\perp\left[\hat{v}_\perp^2 - 2
    \left(1 + \beta_{i }^{-1}\right)\right]g_{\rm 09},
  \label{eq:Gn}
\end{equation}
and
\begin{equation}
  G_B(\bm{r},v_\parallel,t)\; =\;  - \left[\frac{Z}{\tau} + 2\left(1 +
      \beta_{i }^{-1}\right)\right]^{-1} \frac{2\pi}{n_i}
  \int_0^\infty \mathrm{d}v_\perp\, v_\perp\left(\hat{v}_\perp^2  + \frac{Z}{\tau}\right)g_{\rm 09}.
  \label{eq:GB}
\end{equation} 
They then used~(\ref{eq:KRMHDdgdt}) to show that, in the collisionless limit,
\begin{equation}
  \frac{\mathrm{d}G^\pm}{\mathrm{d}t} + v_\parallel \nabla_\parallel G^\pm = \frac{v_\parallel F_{\rm M}(v_\parallel)}{\Lambda^\pm}
\nabla_\parallel \int_{-\infty}^\infty
  \mathrm{d}v_\parallel^\prime G^\pm(\bm{r},v_\parallel^\prime,t),
  \label{eq:dGpmdt}
\end{equation}
where $\mathrm{d}/\mathrm{d}t$ is defined in~(\ref{eq:ddt}), $F_M(v_\parallel) \equiv \pi^{-1/2} v_{\rm th \it i}^{-1} e^{-\hat{v}_\parallel^2}$,
\begin{equation}
  G^+ = G_B + \sigma^{-1}\left(1 + \frac{Z}{\tau}\right) G_n, \qquad
  G^- = G_n + \frac{2\tau}{\sigma Z \beta_{i }}G_B,
  \label{eq:GpGm}
\end{equation}
\begin{equation}
\sigma= 1 + \frac{\tau}{Z} + \beta_{i }^{-1} + \sqrt{\left(1 +
    \frac{\tau}{Z}\right)^2 + \beta_{i }^{-2}},
  \label{eq:g09sigma}
\end{equation}
and
\begin{equation}
  \Lambda^\pm= - \frac{\tau}{Z} + \beta_{i }^{-1} \pm \sqrt{\left(1
      + \frac{\tau}{Z}\right)^2 + \beta_{i }^{-2}}.
  \label{eq:Lambdapm}
\end{equation}
They also employed a Laguerre expansion of~$g$, in which
\begin{equation}
  g(\bm{r},x,v_\parallel,t) = \frac{n_i}{\pi v_{\rm th\it i}^2}e^{-x}\hat{g}(\bm{r},x,v_\parallel,t),
\qquad
\hat{g}(\bm{r},x,v_\parallel,t) = \sum_{l=0}^\infty L_l(x) G_l(\bm{r},v_\parallel,t),
\label{eq:Gl}
\end{equation}
$x= \hat{v}_\perp^2$, $L_l(x) =
(e^x/l!)(\mathrm{d}^l/\mathrm{d}x^l)x^le^{-x}$ is the~$l^{\rm th}$ Laguerre
polynomial, and
\begin{equation}
  G_l(\bm{r}, v_\parallel,t) = \int_0^\infty \mathrm{d}x e^{-x} L_l(x)
  \hat{g}(\bm{r}, x, v_\parallel,t).
  \label{eq:formulaGl}
\end{equation}
It follows from~(\ref{eq:KRMHDdgdt}) that, for $l>1$, 
\begin{equation}
  \frac{\mathrm{d}G_l}{\mathrm{d}t} + v_\parallel \nabla_\parallel G_l = 0.
  \label{eq:dGldt}
\end{equation}
\cite{schekochihin09} defined $\tilde{g}$ to be the part of~$g$ resulting from all
Laguerre moments with~$l\geq 2$:
\begin{equation}
  \tilde{g}(\bm{r},x,v_\parallel,t) = \frac{n_i}{\pi v_{\rm th \it i}^2}e^{-x}\sum_{l=2}^\infty L_l(x) G_l(\bm{r},v_\parallel,t).
    \label{eq:defgtilde}
\end{equation}
Equation~(\ref{eq:dGldt}) implies that
$\mathrm{d}\tilde{g}/\mathrm{d}t + v_\parallel \nabla_\parallel
\tilde{g}= 0.$

\cite{schekochihin09} showed, on the basis of (\ref{eq:dGpmdt})
and~(\ref{eq:dGldt}), that
\begin{equation}
  \frac{\mathrm{d}}{\mathrm{d}t} W^\pm_{\rm compr} = 0, \qquad
  \frac{\mathrm{d}}{\mathrm{d}t} \tilde{W}_{\rm compr} = 0,
  \label{eq:Wpmcons}
\end{equation}
where
\begin{equation}
  W^\pm_{\rm compr} = \frac{n T_{i }}{2}\int \frac{\mathrm{d}^3\bm{r}}{V} 
  \left[ \int_{-\infty}^\infty \frac{\left(G^\pm\right)^2}{F_M}  \mathrm{d}v_\parallel-
    \frac{1}{\Lambda^\pm}\left(\int_{-\infty}^\infty
      \mathrm{d}v_\parallel G^\pm\right)^2\right], 
  \label{eq:Wpmcompr}
\end{equation}
and
\begin{equation}
  \tilde{W}_{\rm compr} = \int \frac{\mathrm{d}^3\bm{r}}{V} \int
  \mathrm{d}^3\bm{v} \frac{T_i \tilde{g}^2}{2 F_{0i}}.
  \label{eq:defWtilde}
\end{equation}
They also showed that the compressive component of the KRMHD free
energy given in~(\ref{eq:KRMHDWcompr}) can be written as\footnote{The coefficients of~$W^\pm_{\rm compr}$ in (\ref{eq:Wcomprtotal})  are
an inconsequential factor of~$1/4$ times
the corresponding coefficients in (213) of~\cite{schekochihin09},
because, when we repeat the calculation of~\cite{schekochihin09}, we
find an extra factor of~1/2 on the
right-hand sides of their~(202) and~(203), which propagates through
(208) and (209) to~(213).}

\begin{align}
  W_{\rm compr} = \tilde{W}_{\rm compr} \,+\,
    \left[ 1 + \frac{1}{\kappa}\left(1 + \frac{\tau}{Z}\right)\right]
                    \left(\Lambda^+\right)^2 W^+_{\rm compr}
 +
      \frac{Z^2}{2\tau^2}\left(1 + \frac{1}{\kappa \beta_{\rm
            i}}\right) \left(\Lambda^-\right)^2 W^-_{\rm
        compr},
    \label{eq:Wcomprtotal}
\end{align}
where
\begin{equation}
  \kappa = \left[ \left(1 + \frac{\tau}{Z}\right) +
    \frac{1}{\beta_i^2}\right]^{1/2}.
  \label{eq:defkappa}
\end{equation}
  
Following a similar set of steps, we find that the phase-space helicity
in KRMHD can be written as the sum of three independently conserved
parts,
\begin{equation}
  H_{\rm \textrm{ph-sp} } = \tilde{H}_{\rm \textrm{ph-sp} } + 
    \left[ 1 + \frac{1}{\kappa}\left(1 + \frac{\tau}{Z}\right)\right]
                    \left(\Lambda^+\right)^2 H^+_{\rm \textrm{ph-sp} }
 +
      \frac{Z^2}{2\tau^2}\left(1 + \frac{1}{\kappa \beta_{\rm
            i}}\right) \left(\Lambda^-\right)^2 H^-_{\rm
        \textrm{ph-sp} },
    \label{eq:Hentrtotal}
\end{equation} 
where
\begin{equation}
  H^\pm_{\rm \textrm{ph-sp} } = n_i T_i \int \frac{\bm{d}^3\bm{r}}{V} \mathrm{P}\int_{-\infty}^\infty \mathrm{d}v_\parallel
\frac{\left(G^\pm\right)^2}{2v_\parallel F_{\rm M}}
\label{eq:Hpm}
\end{equation}
and
\begin{equation}
  \tilde{H}_{\rm \textrm{ph-sp} } = n_i T_i \int \frac{\bm{d}^3\bm{r}}{V} \mathrm{P}\int_{-\infty}^\infty \mathrm{d}v_\parallel\sum_{l=2}^\infty
\frac{G_l^2}{2v_\parallel F_{\rm M}}.
\label{eq:Hentrtilde}
\end{equation}

We also find two additional quadratic invariants.
Equation~(\ref{eq:dGpmdt}) implies that
\begin{align}
  \frac{\mathrm{d}}{\mathrm{d}t} \int\frac{\mathrm{d}^3 \bm{r}}{V} \int_{-\infty}^\infty
  \mathrm{d}v_\parallel \frac{(G^\pm)^2 v_\parallel}{2F_M} & =
                                                             \int\frac{\mathrm{d}^3
                                                             \bm{r}}{V}
                                                             \int_{-\infty}^\infty
                                                             \mathrm{d}v_\parallel
                                                             \frac{G^\pm
                                                             v_\parallel^2}{\Lambda^\pm}
                                                             \nabla_\parallel
                                                             \int_{-\infty}^\infty
                                                             \mathrm{d}v_\parallel^\prime
                                                             G^\pm(\bm{r},
                                                             v_\parallel^\prime,t)
  \label{eq:Khel0} \\
  & = - \frac{1}{\Lambda^\pm} \int\frac{\mathrm{d}^3\bm{r}}{V} M_0^\pm
    \nabla_\parallel M_2^\pm,
    \label{eq:Khel1}
\end{align}
where
\begin{equation}
  M_n^\pm(\bm{r},t)   = \int_{-\infty}^\infty \mathrm{d}v_\parallel v_\parallel^n
  G^\pm(\bm{r},v_\parallel,t),
  \label{eq:Mn}
\end{equation}
and we have integrated by parts to obtain~(\ref{eq:Khel1}) from~(\ref{eq:Khel0}).
Integrating~(\ref{eq:dGpmdt}) over~$v_\parallel$ yields
\begin{equation}
  \frac{\mathrm{d}}{\mathrm{d}t} M_0^\pm  = - \nabla_\parallel M_1^\pm.
  \label{eq:dM0dt}
\end{equation}
Multiplying~(\ref{eq:dGpmdt}) by~$v_\parallel$ and integrating
over~$v_\parallel$ yields
\begin{equation}
  \frac{\mathrm{d}}{\mathrm{d}t} M_1^\pm = - \nabla_\parallel
  M_2^\pm + \frac{v_{\rm th \it i}^2}{2\Lambda^\pm}\nabla_\parallel M_0^\pm.
  \label{eq:dM1dt}
\end{equation}
It then follows from (\ref{eq:Khel1}), (\ref{eq:dM0dt}),
and~(\ref{eq:dM1dt})  that 
\begin{equation}
  \Gamma^\pm \equiv \int\frac{\mathrm{d}^3\bm{r}}{V} \left[
    \int_{-\infty}^\infty \frac{\left(G^\pm\right)^2 v_\parallel}{2 F_{\rm
        M}}\mathrm{d}v_\parallel - \frac{M_0^\pm M_1^\pm}{\Lambda^\pm}\right]
  \label{eq:Gammapm}
\end{equation}
satisfies
\begin{equation}
  \frac{\mathrm{d}\Gamma^\pm}{\mathrm{d}t} =0.
  \label{eq:Gammacons}
\end{equation}

For completeness, we expand~$G^\pm$ in Hermite polynomials,
setting
\begin{equation}
  G^\pm(\bm{r}, v_\parallel, t) = \sum_{m=0}^\infty
  \frac{H_m(\hat{v}_\parallel) F_M(v_\parallel)}{\sqrt{2^m
      m!}}\tilde{G}^\pm_m(\bm{r},t),
  \label{eq:hermiteexp}
\end{equation}
where $H_m(x) = (-1)^m e^{x^2}(\mathrm{d}^m/\mathrm{d}x^m) e^{-x^2}$ is
the Hermite polynomial of order~$m$, which satisfies the recurrence
relation~
\begin{equation}
  \hat{v}_\parallel
H_m(\hat{v}_\parallel) = \frac{1}{2} H_{m+1}(\hat{v}_\parallel) + m
H_{m-1}(\hat{v}_\parallel).
\label{eq:recurrence} 
\end{equation}
Multiplying
(\ref{eq:hermiteexp}) by~$H_n(\hat{v}_\parallel)$, integrating
over~$v_\parallel$, and applying the orthogonality relation
\begin{equation}
  \int_{-\infty}^\infty \mathrm{d}v_\parallel H_m(\hat{v}_\parallel)
  H_n(\hat{v}_\parallel) F_M(v_\parallel) = 2^m m! \,\delta_{mn}
  \label{eq:orthogonality}
\end{equation}
yields
\begin{equation}
\tilde{G}^\pm_n(\bm{r},t) = \int_{-\infty}^\infty \mathrm{d}v_\parallel
\frac{H_n(\hat{v}_\parallel)}{\sqrt{2^n n!}}G^\pm(\bm{r},v_\parallel,
t).
\label{eq:valGm}
\end{equation}
Upon substituting these expressions into (\ref{eq:Wpmcompr})
and~(\ref{eq:Gammapm}), we obtain
\begin{equation}
  W_{\rm compr}^\pm = \frac{n T_i}{2} \int \frac{\mathrm{d}^3
    \bm{r}}{V}\left[ \sum_{m=0}^\infty \left(\tilde{G}_m^\pm\right)^2 -
    \frac{\left( \tilde{G}^\pm_0 \right)^2}{\Lambda^\pm}\right]
  \label{eq:WcomprHerm} 
\end{equation}
and
\begin{equation}
  \Gamma^\pm = \frac{v_{\rm th\it i}}{\sqrt{2}}\int \frac{\mathrm{d}^3\bm{r}}{V} \left(
\sum_{m=0}^\infty \tilde{G}^\pm_m \tilde{G}^\pm_{m+1}
\sqrt{m+1} - \frac{\tilde{G}^\pm_0 \tilde{G}^\pm_1}{\Lambda^\pm}
    \right).
    \label{eq:GammapmHerm}
\end{equation}
If one were to view phase mixing as a local `cascade' towards
larger~$m$ (albeit one mediated by the linear
$v_\parallel \partial g_{09}/\partial z$ term in~(\ref{eq:KRMHDdgdt})
rather than the nonlinear terms), and if one were to approximate
$|\tilde{G}^\pm_m \tilde{G}^\pm_{m+1} \sqrt{m+1}| \simeq |m^{1/2}
(\tilde{G}^\pm_m)^2|$ in~(\ref{eq:GammapmHerm}), then it might at
first appear that the factor of~$\sqrt{m+1}$ in~(\ref{eq:GammapmHerm})
causes~$W^\pm_{\rm compr}$ and~$\Gamma^\pm$ to satisfy~(\ref{eq:AB1}),
with the invariant~A in~(\ref{eq:AB1}) corresponding
to~$W^\pm_{\rm compr}$, and the invariant~B corresponding
to~$\Gamma^\pm$. If~(\ref{eq:AB1}) were satisfied in this way,
then~$W^\pm_{\rm compr}$ would be unable to `cascade' to larger~$m$.
However, $\Gamma^\pm$ is not sign-definite, because~$\tilde{G}^\pm_{m}$
and~$\tilde{G}^\pm_{m+1}$ need not have the same sign. Because of this,
$W^\pm_{\rm compr}$ and~$\Gamma^\pm$ do not satisfy~(\ref{eq:AB1}),
and we expect both to cascade to larger~$m$, as discussed at the end
of~\S\ref{sec:Fjortoft}.

\bibliography{articles}

\begin{thebibliography}{29}
\expandafter\ifx\csname natexlab\endcsname\relax\def\natexlab#1{#1}\fi
\def\au#1{#1} \def\ed#1{#1} \def\yr#1{#1}\def\at#1{#1}\def\jt#1{\textit{#1}}
  \def\bt#1{#1}\def\bvol#1{\textbf{#1}} \def\vol#1{#1} \def\pg#1{#1}
  \def\publ#1{#1}\def\arxiv#1{#1}\def\org#1{#1}\def\st#1{\textit{#1}}

\bibitem[{Abel} {\em et~al.\/}(2013){Abel}, {Plunk}, {Wang}, {Barnes},
  {Cowley}, {Dorland} \& {Schekochihin}]{abel13}
{\sc \au{{Abel}, I.~G.}, \au{{Plunk}, G.~G.}, \au{{Wang}, E.}, \au{{Barnes},
  M.}, \au{{Cowley}, S.~C.}, \au{{Dorland}, W.} \& \au{{Schekochihin}, A.~A.}}
  \yr{2013}  \at{{Multiscale gyrokinetics for rotating tokamak plasmas:
  fluctuations, transport and energy flows}}.  \jt{Reports on Progress in
  Physics}  \bvol{76}~(11),  \pg{116201},  \arxiv{arXiv: 1209.4782}.

\bibitem[{Adkins} {\em et~al.\/}(2024){Adkins}, {Meyrand} \&
  {Squire}]{adkins24}
{\sc \au{{Adkins}, T.}, \au{{Meyrand}, R.} \& \au{{Squire}, J.}} \yr{2024}
  \at{{The effects of finite electron inertia on helicity-barrier-mediated
  turbulence}}.  \jt{Journal of Plasma Physics}  \bvol{90}~(4),
  \pg{905900403},  \arxiv{arXiv: 2404.09380}.

\bibitem[{Adkins} {\em et~al.\/}(2025){Adkins}, {Meyrand} \&
  {Squire}]{adkins25}
{\sc \au{{Adkins}, T.}, \au{{Meyrand}, R.} \& \au{{Squire}, J.}} \yr{2025}
  \at{{Turbulent Heating in Collisionless Low-beta Plasmas: Imbalance, Landau
  Damping, and Electron-Ion Energy Partition}}.  \jt{\apj}  \bvol{990}~(2),
  \pg{138},  \arxiv{arXiv: 2504.16177}.

\bibitem[{Alexakis} \& {Biferale}(2018)]{alexakis18}
{\sc \au{{Alexakis}, A.} \& \au{{Biferale}, L.}} \yr{2018}  \at{{Cascades and
  transitions in turbulent flows}}.  \jt{\physrep}  \bvol{767},  \pg{1--101},
  \arxiv{arXiv: 1808.06186}.

\bibitem[{Antonsen} \& {Lane}(1980)]{antonsen80}
{\sc \au{{Antonsen}, Jr., T.~M.} \& \au{{Lane}, B.}} \yr{1980}  \at{{Kinetic
  equations for low frequency instabilities in inhomogeneous plasmas}}.
  \jt{Physics of Fluids}  \bvol{23}~(6),  \pg{1205--1214}.

\bibitem[{Catto}(1978)]{catto78}
{\sc \au{{Catto}, P.~J.}} \yr{1978}  \at{{RESEARCH NOTES: Linearized
  gyro-kinetics}}.  \jt{Plasma Physics}  \bvol{20}~(7),  \pg{719--722}.

\bibitem[{Coles} \& {Harmon}(1989)]{coles89}
{\sc \au{{Coles}, W.~A.} \& \au{{Harmon}, J.~K.}} \yr{1989}  \at{{Propagation
  observations of the solar wind near the sun}}.  \jt{\apj}  \bvol{337},
  \pg{1023--1034}.

\bibitem[{Cranmer} \& {van Ballegooijen}(2005)]{cranmer05}
{\sc \au{{Cranmer}, S.~R.} \& \au{{van Ballegooijen}, A.~A.}} \yr{2005}  \at{On
  the generation, propagation, and reflection of {A}lfv{\' e}n waves from the
  solar photosphere to the distant heliosphere}.  \jt{\apjs}  \bvol{156},
  \pg{265--293}.

\bibitem[{Cuesta} {\em et~al.\/}(2023){Cuesta}, {Chhiber}, {Fu}, {Du}, {Yang},
  {Pecora}, {Matthaeus}, {Li}, {Steinberg}, {Guo}, {Gan}, {Conrad} \&
  {Swanson}]{cuesta23}
{\sc \au{{Cuesta}, M.~E.}, \au{{Chhiber}, R.}, \au{{Fu}, X.}, \au{{Du}, S.},
  \au{{Yang}, Y.}, \au{{Pecora}, F.}, \au{{Matthaeus}, W.~H.}, \au{{Li}, H.},
  \au{{Steinberg}, J.}, \au{{Guo}, F.}, \au{{Gan}, Z.}, \au{{Conrad}, E.} \&
  \au{{Swanson}, D.}} \yr{2023}  \at{{Compressible Turbulence in the Near-Sun
  Solar Wind: Parker Solar Probe's First Eight Perihelia}}.  \jt{\apjl}
  \bvol{949}~(2),  \pg{L19},  \arxiv{arXiv: 2305.03566}.

\bibitem[{De Pontieu} {\em et~al.\/}(2007){De Pontieu}, {McIntosh}, {Carlsson},
  {Hansteen}, {Tarbell}, {Schrijver}, {Title}, {Shine}, {Tsuneta}, {Katsukawa},
  {Ichimoto}, {Suematsu}, {Shimizu} \& {Nagata}]{depontieu07}
{\sc \au{{De Pontieu}, B.}, \au{{McIntosh}, S.~W.}, \au{{Carlsson}, M.},
  \au{{Hansteen}, V.~H.}, \au{{Tarbell}, T.~D.}, \au{{Schrijver}, C.~J.},
  \au{{Title}, A.~M.}, \au{{Shine}, R.~A.}, \au{{Tsuneta}, S.},
  \au{{Katsukawa}, Y.}, \au{{Ichimoto}, K.}, \au{{Suematsu}, Y.},
  \au{{Shimizu}, T.} \& \au{{Nagata}, S.}} \yr{2007}  \at{{Chromospheric
  Alfv{\'e}nic Waves Strong Enough to Power the Solar Wind}}.  \jt{Science}
  \bvol{318},  \pg{1574--7}.

\bibitem[{Dmitruk} {\em et~al.\/}(2002){Dmitruk}, {Matthaeus}, {Milano},
  {Oughton}, {Zank} \& {Mullan}]{dmitruk02}
{\sc \au{{Dmitruk}, P.}, \au{{Matthaeus}, W.~H.}, \au{{Milano}, L.~J.},
  \au{{Oughton}, S.}, \au{{Zank}, G.~P.} \& \au{{Mullan}, D.~J.}} \yr{2002}
  \at{Coronal heating distribution due to low-frequency, wave-driven
  turbulence}.  \jt{\apj}  \bvol{575},  \pg{571--577}.

\bibitem[{Fj{\o}rtoft}(1953)]{fjortoft53}
{\sc \au{{Fj{\o}rtoft}, R.}} \yr{1953}  \at{{On the Changes in the Spectral
  Distribution of Kinetic Energy for Twodimensional, Nondivergent Flow}}.
  \jt{Tellus}  \bvol{5}~(3),  \pg{225--230}.

\bibitem[{Frieman} \& {Chen}(1982)]{frieman82}
{\sc \au{{Frieman}, E.~A.} \& \au{{Chen}, L.}} \yr{1982}  \at{{Nonlinear
  gyrokinetic equations for low-frequency electromagnetic waves in general
  plasma equilibria}}.  \jt{Physics of Fluids}  \bvol{25}~(3),  \pg{502--508}.

\bibitem[{Harmon} \& {Coles}(2005)]{harmon05}
{\sc \au{{Harmon}, J.~K.} \& \au{{Coles}, W.~A.}} \yr{2005}  \at{{Modeling
  radio scattering and scintillation observations of the inner solar wind using
  oblique Alfv{\'e}n/ion cyclotron waves}}.  \jt{Journal of Geophysical
  Research (Space Physics)}  \bvol{110},  \pg{3101--+}.

\bibitem[{Howes} {\em et~al.\/}(2006){Howes}, {Cowley}, {Dorland}, {Hammett},
  {Quataert} \& {Schekochihin}]{howes06}
{\sc \au{{Howes}, G.~G.}, \au{{Cowley}, S.~C.}, \au{{Dorland}, W.},
  \au{{Hammett}, G.~W.}, \au{{Quataert}, E.} \& \au{{Schekochihin}, A.~A.}}
  \yr{2006}  \at{{Astrophysical Gyrokinetics: Basic Equations and Linear
  Theory}}.  \jt{\apj}  \bvol{651}~(1),  \pg{590--614},  \arxiv{arXiv:
  astro-ph/0511812}.

\bibitem[{Kawazura} \& {Barnes}(2018)]{kawazura18}
{\sc \au{{Kawazura}, Y.} \& \au{{Barnes}, M.}} \yr{2018}  \at{{A hybrid
  gyrokinetic ion and isothermal electron fluid code for astrophysical
  plasma}}.  \jt{Journal of Computational Physics}  \bvol{360},  \pg{57--73},
  \arxiv{arXiv: 1708.07235}.

\bibitem[{Kawazura} {\em et~al.\/}(2019){Kawazura}, {Barnes} \&
  {Schekochihin}]{kawazura19}
{\sc \au{{Kawazura}, Y.}, \au{{Barnes}, M.} \& \au{{Schekochihin}, A.~A.}}
  \yr{2019}  \at{{Thermal disequilibration of ions and electrons by
  collisionless plasma turbulence}}.  \jt{\pnas}  \bvol{116}~(3),
  \pg{771--776},  \arxiv{arXiv: 1807.07702}.

\bibitem[{Meyrand} {\em et~al.\/}(2021){Meyrand}, {Squire}, {Schekochihin} \&
  {Dorland}]{meyrand21}
{\sc \au{{Meyrand}, R.}, \au{{Squire}, J.}, \au{{Schekochihin}, A.~A.} \&
  \au{{Dorland}, W.}} \yr{2021}  \at{{On the violation of the zeroth law of
  turbulence in space plasmas}}.  \jt{Journal of Plasma Physics}
  \bvol{87}~(3),  \pg{535870301},  \arxiv{arXiv: 2009.02828}.

\bibitem[{Milanese} {\em et~al.\/}(2021){Milanese}, {Loureiro} \&
  {Boldyrev}]{milanese21}
{\sc \au{{Milanese}, L.~M.}, \au{{Loureiro}, N.~F.} \& \au{{Boldyrev}, S.}}
  \yr{2021}  \at{{Dynamic Phase Alignment in Navier-Stokes Turbulence}}.
  \jt{\prl}  \bvol{127}~(27),  \pg{274501},  \arxiv{arXiv: 2104.13518}.

\bibitem[{Raymond} {\em et~al.\/}(2014){Raymond}, {McCauley}, {Cranmer} \&
  {Downs}]{raymond14}
{\sc \au{{Raymond}, J.~C.}, \au{{McCauley}, P.~I.}, \au{{Cranmer}, S.~R.} \&
  \au{{Downs}, C.}} \yr{2014}  \at{{The Solar Corona as Probed by Comet Lovejoy
  (C/2011 W3)}}.  \jt{\apj}  \bvol{788},  \pg{152},  \arxiv{arXiv: 1405.1639}.

\bibitem[{Rutherford} \& {Frieman}(1968)]{rutherford68}
{\sc \au{{Rutherford}, P.~H.} \& \au{{Frieman}, E.~A.}} \yr{1968}  \at{{Drift
  Instabilities in General Magnetic Field Configurations}}.  \jt{Physics of
  Fluids}  \bvol{11}~(3),  \pg{569--585}.

\bibitem[{Schekochihin} {\em et~al.\/}(2009){Schekochihin}, {Cowley},
  {Dorland}, {Hammett}, {Howes}, {Quataert} \& {Tatsuno}]{schekochihin09}
{\sc \au{{Schekochihin}, A.~A.}, \au{{Cowley}, S.~C.}, \au{{Dorland}, W.},
  \au{{Hammett}, G.~W.}, \au{{Howes}, G.~G.}, \au{{Quataert}, E.} \&
  \au{{Tatsuno}, T.}} \yr{2009}  \at{{Astrophysical Gyrokinetics: Kinetic and
  Fluid Turbulent Cascades in Magnetized Weakly Collisional Plasmas}}.
  \jt{\apjs}  \bvol{182},  \pg{310--377},  \arxiv{arXiv: 0704.0044}.

\bibitem[{Schekochihin} {\em et~al.\/}(2019){Schekochihin}, {Kawazura} \&
  {Barnes}]{schekochihin19}
{\sc \au{{Schekochihin}, A.~A.}, \au{{Kawazura}, Y.} \& \au{{Barnes}, M.~A.}}
  \yr{2019}  \at{{Constraints on ion versus electron heating by plasma
  turbulence at low beta}}.  \jt{Journal of Plasma Physics}  \bvol{85}~(3),
  \pg{905850303},  \arxiv{arXiv: 1812.09792}.

\bibitem[{Squire} {\em et~al.\/}(2023){Squire}, {Meyrand} \& {Kunz}]{squire23}
{\sc \au{{Squire}, J.}, \au{{Meyrand}, R.} \& \au{{Kunz}, M.~W.}} \yr{2023}
  \at{{Electron-Ion Heating Partition in Imbalanced Solar-wind Turbulence}}.
  \jt{\apjl}  \bvol{957}~(2),  \pg{L30},  \arxiv{arXiv: 2308.13048}.

\bibitem[{Squire} {\em et~al.\/}(2022){Squire}, {Meyrand}, {Kunz},
  {Arzamasskiy}, {Schekochihin} \& {Quataert}]{squire22}
{\sc \au{{Squire}, J.}, \au{{Meyrand}, R.}, \au{{Kunz}, M.~W.},
  \au{{Arzamasskiy}, L.}, \au{{Schekochihin}, A.~A.} \& \au{{Quataert}, E.}}
  \yr{2022}  \at{{High-frequency heating of the solar wind triggered by
  low-frequency turbulence}}.  \jt{Nature Astronomy}  \bvol{6},  \pg{715--723},
   \arxiv{arXiv: 2109.03255}.

\bibitem[{Taylor} \& {Hastie}(1968)]{taylor68}
{\sc \au{{Taylor}, J.~B.} \& \au{{Hastie}, R.~J.}} \yr{1968}  \at{{Stability of
  general plasma equilibria - I formal theory}}.  \jt{Plasma Physics}
  \bvol{10}~(5),  \pg{479--494}.

\bibitem[{Verdini} \& {Velli}(2007)]{verdini07}
{\sc \au{{Verdini}, A.} \& \au{{Velli}, M.}} \yr{2007}  \at{{A}lfv{\'e}n waves
  and turbulence in the solar atmosphere and solar wind}.  \jt{\apj}
  \bvol{662},  \pg{669--676},  \arxiv{arXiv: arXiv:astro-ph/0702205}.

\bibitem[{Zhang} {\em et~al.\/}(2025){Zhang}, {Kunz}, {Squire} \&
  {Klein}]{zhang25}
{\sc \au{{Zhang}, M.~F.}, \au{{Kunz}, M.~W.}, \au{{Squire}, J.} \& \au{{Klein},
  K.~G.}} \yr{2025}  \at{{Extreme Heating of Minor Ions in Imbalanced
  Solar-wind Turbulence}}.  \jt{\apj}  \bvol{979}~(2),  \pg{121},
  \arxiv{arXiv: 2408.04703}.

\bibitem[{Zocco} \& {Schekochihin}(2011)]{zocco11}
{\sc \au{{Zocco}, A.} \& \au{{Schekochihin}, A.~A.}} \yr{2011}  \at{{Reduced
  fluid-kinetic equations for low-frequency dynamics, magnetic reconnection,
  and electron heating in low-beta plasmas}}.  \jt{Physics of Plasmas}
  \bvol{18}~(10),  \pg{102309--102309},  \arxiv{arXiv: 1104.4622}.

\end{thebibliography}

\bibliographystyle{jpp}

\end{document}